\documentclass[12pt]{article}

\usepackage{psfrag,epsf}
\usepackage{enumerate}
\usepackage{url}

\usepackage[english]{babel}
\usepackage{csquotes}
\usepackage{array}
\usepackage{float}
\usepackage{amsmath}
\usepackage{amsfonts}
\usepackage{lscape}
\usepackage{amssymb}
\usepackage{latexsym}
\usepackage{epsfig}
\usepackage{graphicx}
\usepackage{mathrsfs}
\usepackage{fancyhdr}
\usepackage{sectsty}
\usepackage{caption}
\usepackage{setspace}
\usepackage[colorlinks]{hyperref}
\usepackage{arydshln}
\usepackage{xcolor}
\hypersetup{
	colorlinks,
	linkcolor={red!50!black},
	citecolor={green!30!black},
	urlcolor={blue!80!black}
}
\usepackage{theorem}
\usepackage{url}
\usepackage{rotating}
\usepackage{multirow}
\usepackage{tabularx}
\usepackage{graphics}
\usepackage{float}
\usepackage{afterpage}
\usepackage{booktabs}
{
\theorembodyfont{\upshape} 

\theoremheaderfont{\scshape} 

}
\newtheorem{assumption}{Assumption}

{\theorembodyfont{\upshape} 

}

\newtheorem{proposition}{{\sc Proposition}}

\newenvironment{proof}[1][Proof]{\bigskip \noindent \textbf{#1:} }{\  \rule{0.5em}{0.5em}}

\theoremheaderfont{\bfseries}
\theorembodyfont{\upshape}

\sectionfont{\centering \sc}
\subsectionfont{\centering \sc}
\subsubsectionfont{\centering \sc}
\allsectionsfont{\centering \sc}
\allowdisplaybreaks
\renewcommand{\baselinestretch}{1.1}
\renewcommand{\baselinestretch}{1.05}

\usepackage[round,authoryear]{natbib}   
\bibpunct{(}{)}{;}{a}{,}{,}            

\hypersetup{
	pdfborderstyle={/S/U/W 1}, 
}

\renewcommand{\cite}{\citet}

\begin{document}
\def\spacingset#1{\renewcommand{\baselinestretch}{#1}\small\normalsize} \spacingset{1}

\begin{center}
\bigskip {\Large{\textsc{Invalid Proxies and Volatility Changes \\ }}}
\renewcommand{\thefootnote}{}
\footnote{
\hspace{-7.2mm}
$^{a}$Department of Economics, University of Bologna, Italy.
$^{b}$CORE, UCLouvain. 
Correspondence to: Luca Fanelli, Department of Economics, University of Bologna, Piazza Scaravilli 2, 
40126 Bologna, Italy; email: luca.fanelli@unibo.it. 
}
\addtocounter{footnote}{-1}
\renewcommand{\thefootnote}{\arabic{footnote}}
{\normalsize \vspace{0.1cm} }
\thispagestyle{empty}
\textsc{Giovanni Angelini}$^{a}$\textsc{, Luca Fanelli}$^{a}$\textsc{, Luca Neri}$^{b}$ \\
\addtocounter{page}{1}
\medskip
This version: \today
\end{center}
\par
\begingroup
\leftskip = 1 cm
\rightskip = 1 cm
\small

\begin{center}
\textsc{Abstract}
\end{center}

When in proxy-SVARs the covariance matrix of VAR disturbances is subject to exogenous, permanent breaks that cause IRFs to change across volatility regimes, even strong, exogenous external instruments yield inconsistent estimates of the dynamic causal effects. However, if these volatility shifts are properly incorporated into the analysis through (testable) ``stability restrictions'', we demonstrate that the target IRFs are point-identified and can be estimated consistently under a necessary and sufficient rank condition. If the shifts in volatility are sufficiently informative, standard asymptotic inference remains valid even with (i) local-to-zero covariance between the proxies and the instrumented structural shocks, and (ii) potential failures of instrument exogeneity. Intuitively, shifts in volatility act similarly to strong instruments that are correlated with both the target and non-target shocks. We illustrate the effectiveness of our approach by revisiting a seminal fiscal proxy-SVAR for the US economy. We detect a sharp change in the size of the tax multiplier when the narrative tax instrument is complemented with the decline in unconditional volatility observed during the transition from the Great Inflation to the Great Moderation. The narrative tax instrument contributes to identify the tax shock in both regimes, although our empirical analysis raises concerns about its statistical validity.

\bigskip


\bigskip


\vspace{0.15cm}

\par
\endgroup
\spacingset{1.8} 
\section{Introduction}

Structural Vector Autoregressions (SVARs) identified by external instruments or proxies, hereafter proxy-SVARs, are commonly used alongside or as alternatives to local projections for identifying macroeconomic shocks. Proxy-SVARs address a partial identification problem by incorporating external instruments into the SVAR. Proxies must satisfy two key conditions: relevance (correlation with target shocks) and exogeneity (no correlation with non-target shocks); see 
\cite{MertensRavn2013} and \cite{StockWatson2018}. When both relevance and exogeneity conditions hold, the target impulse response functions (IRFs) are point-identified, and their estimator is consistent and asymptotically normal. \cite{MontielOleaStockWatson2021} extend asymptotic inference in proxy-SVARs to scenarios where proxies are \textquotedblleft weak\textquotedblright\ as defined in \cite{StaigerStock1997}. Their work highlights that even external variables weakly correlated with the target structural shocks possess valuable information for identification, although inference may become conservative in finite samples. We show that under certain conditions, even weak and possibly contaminated (i.e., correlated with non-target shocks) instruments may possess valuable information for identification.

Economic relationships are often subject to structural breaks, typically caused by shifts in underlying behavior, market conditions, or policy conduct. These breaks are pervasive in macroeconomics, and proxy-SVARs are not immune. The conventional ``identification-through-heteroskedasticity" approach, extended by 
\cite{LanneLutkepohl2008} to SVARs, assumes constant IRFs across volatility regimes, implying that the impact and propagation of structural shocks remain constant, up to scale, across different macroeconomic regimes. This limits the scope and potential of proxy-SVAR analysis. We show that volatility breaks that induce changes in IRFs tend to compromise the consistency of estimators  even when strong and exogenous instruments are used. A common alternative approach, split-sampling, requires estimating separate models before and after the break, overlooking valuable information for identification. Splitting the sample can reduce estimation precision (as the effective sample size is reduced) and can make external instruments appear weaker than they truly are \citep[see][]{antoine2024efficient}.

We demonstrate that external instruments remain valuable for identifying target shocks in the presence of volatility breaks that change IRFs, provided that (i) the moment conditions implied by the volatility shifts are properly incorporated into the proxy-SVAR, and (ii) the volatility breaks are sufficiently informative. The latter requirement translates into a necessary and sufficient rank condition that allows instruments to contribute to the identification process even when they are weak or contaminated. Notably, in the worst case scenarios, economically significant but statistically invalid instruments can still serve as labels for the structural shocks. 

The flexibility of the proposed approach is empirically relevant. In macroeconomics, proxies are often weak and may display substantial variability in their strength: there are periods or events where they are very informative about the shocks of interest, followed by periods where they become poorly informative. Moreover, the low-frequency nature of macroeconomic data increases the potential for contamination by confounding factors.

Focusing on proxy-SVARs with a finite number of distinct, permanent volatility regimes, we allow IRFs to vary across these regimes. Estimators that ignore volatility shifts are inconsistent in this framework. However, incorporating volatility shifts via theory-driven stability restrictions - i.e., taking a stand on which structural parameters vary across regimes and which remain constant \citep[see][]{MagnussonMavroeidis2014} - allows to point-identify and consistently estimate the target IRFs. Interestingly, stability restrictions encompass the conventional identification-through-heteroskedasticity approach. In addition, stability restrictions often lead to overidentification, thus facilitating specification testing.  By relying on economic reasoning other than the statistical feature of the data, our approach overcomes limitations of purely statistical identification methods emphasized in, e.g., \cite{MontielOleaPlagborg-MollerQian2022}.

When volatility shifts are informative enough to meet our necessary and sufficient rank condition under the specified stability restrictions, the target IRFs are estimated consistently and standard inference applies. This is true even in the presence of weak or contaminated instruments. Intuitively, under the rank condition, shifts in volatility act similarly to strong instruments that are correlated with both the target and non-target shocks. Accordingly, weak instruments do not lead to non-standard asymptotics if volatility shifts compensate with sufficient identification information. Furthermore, they do not lead to efficiency losses. These results align with \cite{AntoineRenault2017}'s findings on the relevance of weak instruments in GMM estimation when also strong instruments are available. Finally, our framework allows relaxing the exogeneity condition since volatility shifts provide information on both target and non-target shocks. This ensures the consistent estimation of target shocks despite nonzero correlations between proxies and non-target shocks. It turns out that external instruments, considered "informative" on the target structural shocks from an economic standpoint, can still be employed in the analysis despite the possible failure of their statistical properties. This has also implications on the identification strategy: in the presence of informative breaks in unconditional volatility, the set of proxies that can be potentially used in empirical analysis increases. For instance, practitioners can focus on the relevance of the proxies selected for the target shocks without being overly concerned about the exogeneity condition.

We apply our stability restrictions approach by augmenting the proxy-SVAR with instrument equations \citep[see e.g.,][]{AngeliniFanelli2019,Ariasetal2021,GiacominiKitagawaRead2022}. Shifts in the error covariance matrix of this enlarged system capture changes in the unconditional volatility of the variables, including potential changes in parameters related to relevance, contamination, and variance of instruments' measurement errors. We introduce a Classical Minimum Distance (CMD) estimation method (and an alternative Quasi Maximum Likelihood (QML) approach in the supplementary material), where identification is ensured by the rank of the Jacobian matrix derived from the mapping between reduced form and structural parameters,  under the specified stability restrictions. Monte Carlo simulations show that combining strong exogenous instruments with volatility shifts enhances estimation precision compared to using volatility shifts alone. Even with contaminated but strongly relevant instruments, precision improves significantly, and there are no precision losses with weak and contaminated instruments. The overidentifying restrictions test effectively detects misspecified stability restrictions.

Our empirical illustration revisits the seminal fiscal proxy-SVAR estimated in \cite{MertensRavn2014} to infer fiscal multipliers, using US quarterly data from 1950:Q1 to 2006:Q4.  We estimate a break in 1983:Q2 and apply our stability restrictions approach to account for the volatility decline from the Great Inflation to the Great Moderation, allowing IRFs to change across these two macroeconomic regimes.  

\paragraph{Connections with the literature}
Our approach relates to and extends existing literature. \cite{SchlaakRiethPodstawski2023} highlight the benefits of volatility breaks for  testing instrument exogeneity in point-identified proxy-SVARs. In their setup, a single instrument is used for a single target shock and IRFs are assumed constant across volatility regimes. Our Proposition \ref{Proposition 1} shows that, in their framework, consistent IRF estimation is achievable even ignoring volatility breaks if instruments are valid. Additionally, similar to \cite{LudvigsonMaNg2021} and \cite{BraunBruggermann2023}, our method does not require imposing instrument exogeneity prior estimation. Compared to \cite{KewelohKleinPruser2024}, who also maintain constant IRFs, our stability restrictions do not assume specific distributions or independence of structural shocks. Unlike \cite{carriero2024blended}, who again assume constant IRFs across regimes, our framework accommodates regime-dependent IRFs by integrating stability restrictions for consistent IRF estimation.  In \cite{carriero2024blended}, the idea is that heteroskedasticity can improve identification and relax the need for strict zero restrictions that are often necessary in proxy-SVARs with multiple target shocks. Our analysis demonstrates that, if the assumption of constant IRFs across volatility regimes is not empirically tenable, estimators of IRFs are inconsistent.
Finally, our work extends  \citet{LutkepohlSchlaak2022} and \citet{BrunsLutkepohl2024} by allowing impulse‑response functions to vary across regimes without assuming instruments that remain valid and time‑invariant.

\paragraph{Structure of the paper}
The paper is organized as follows. Section \ref{Section_two_volatility_regimes} introduces our baseline proxy-SVAR with a volatility break. Section \ref{Model_Setup} presents the augmented SVAR framework and defines instrument properties. Section \ref{Section_Estimation_ignoring_break} examines conditions for consistent IRF estimation when ignoring volatility breaks. Section \ref{Section_stability_restrictions_approach} details the stability restrictions approach, including identification and CMD estimation. Section \ref{Section_MC_main_test} reports part of our Monte Carlo results on the relative performance of our approach and the finite sample properties of the overidentifying restrictions test implied by the CMD estimation approach. Section \ref{Section_empirical_illustration} applies the methodology to US fiscal multipliers. Section \ref{Section_concluding_remarks} concludes. A supplementary material provides additional details, including proofs and extended analyses.

\section{Proxy-SVARs with a Shift in Unconditional Volatility}

\label{Section_two_volatility_regimes}In this section, we introduce our
approach to proxy-SVARs within a DGP that incorporates a single break ($M=1$%
) in the error covariance m atrix, resulting in two ($M+1=2$) volatility
regimes in the data. The one break-two volatility regimes model is presented
for clarity of exposition; the analysis is extended to more than one
structural break in Section S.4 of
the supplementary material.

\subsection{Baseline Proxy-SVAR and Proxy Properties}
\label{Model_Setup}Our baseline is the SVAR model:
{\setlength{\abovedisplayskip}{5pt}
\setlength{\belowdisplayskip}{5pt}
\begin{equation}
\begin{array}{lll}
Y_{t} =\Pi\,  X_{t}+u_{t}, & \quad u_{t} =H\, \varepsilon _{t}, & \quad t=1,...,T  \label{VAR-RF2} 
 \end{array}%
 \end{equation}
 }
where $Y_{t}$ is the $n\times 1$ vector of endogenous variables, $%
X_{t}:=(Y_{t-1}^{\prime },...,Y_{t-p}^{\prime })^{\prime }$ is the vector
collecting $p$ lags of the variables, $T$ is the number of length of the sample, $%
\Pi :=(\Pi _{1},$ $...$ $,\Pi _{p})$ is the $n\times np$ matrix containing
the autoregressive (slope) parameters. Finally, $u_t$ is the $n$-dimensional vector containing the VAR innovations. We assume $u_t$ is a vector of martingale 
difference sequences (MDS), such that $\mathbb{E}(u_t \mid \mathcal{I}_{t-1}) = 0_{n \times 1}$ (a.s.), 
where $\mathcal{I}_t := \sigma(\,Y_t,\, Y_{t-1}, \, \ldots)$ denotes the $\sigma$-algebra 
generated by the information available at time $t$.  
Deterministic terms
have been excluded from Equation \eqref{VAR-RF2} without loss of generality. The initial
values $Y_{0},...,Y_{1-p}$ are treated as fixed constants throughout the
analysis. In what follows, we denote the VAR companion matrix by $\mathcal{C}_{\Pi}$. This matrix depends on the parameters in $\Pi$, specifically, $\mathcal{C}_{\Pi} := \mathcal{C}(\Pi)$, where $\mathcal{C}(\bullet)$ is the matrix-valued function representing the SVAR in its state-space form.

In Equation \eqref{VAR-RF2}, the system of equations $u_{t}=H\,\varepsilon _{t}$\ maps
the $n\times 1$ vector of structural shocks $\varepsilon _{t}$ to the
reduced form innovations through the columns of the $n\times n$ matrix $H$. Matrix $H$ is assumed non-singular and its rows contain the on-impact (instantaneous) effects of the structural shocks onto the endogenous variables. Except where otherwise indicated, the structural shocks have normalized covariance
matrix $\Sigma _{\varepsilon }:=\mathbb{E}(\varepsilon _{t}\varepsilon
_{t}^{\prime })=I_{n}$. Furthermore, we temporarily assume that the reduced form parameters 
$(\Pi ,\, \Sigma _{u})$, are time-invariant over the sample
 $Y_{1},...,Y_{T}$. This assumption will be relaxed in Section \ref%
{Section_DGP}.

Let $\varepsilon _{1,t}$ be the $k\times 1$ sub-vector of elements in $%
\varepsilon _{t}$ containing the $1\leq k\leq n$ target structural
shocks. We consider a corresponding partition of the structural relationship:
{\setlength{\abovedisplayskip}{5pt}
\setlength{\belowdisplayskip}{5pt}
\begin{equation}
u_{t} :=\left( 
\begin{array}{l}
u_{1,t} \\ 
u_{2,t}%
\end{array}%
\right) =\left( 
\begin{array}{cc}
H_{1,1} & H_{1,2} \\ 
H_{2,1} & H_{2,2}%
\end{array}%
\right) \left( 
\begin{array}{l}
\varepsilon _{1,t} \\ 
\varepsilon _{2,t}%
\end{array}%
\right) =H_{\bullet 1}\varepsilon _{1,t}+\text{ }H_{\bullet 2}\varepsilon
_{2,t}  \label{partition_B}
\end{equation}%
}
where $\varepsilon _{2,t}$ contains the $(n-k)$ structural shocks that are
not of interest. VAR disturbances $u_{1,t}$ and $u_{2,t}$  have the
same dimensions as $\varepsilon _{1,t}$ and $\varepsilon _{2,t}$,
respectively. Matrix $H_{\bullet 1}:=(H_{1,1}^{\prime }, \, \, H_{2,1}^{\prime
})^{\prime }$ is of dimension $n\times k$ and collects the on-impact coefficients
associated with the target structural shocks. Finally, $H_{\bullet 2}$ is of dimension $%
n\times (n-k)$\ and collects the on-impact coefficients associated with the
non-target shocks.

The objective of the analysis is to identify and estimate the $\ell$-period-ahead 
responses of the variables $Y_{t+\ell}$ to the $j$-th target shock in $\varepsilon_{1,t}$. 
These responses are given by: 
{\setlength{\abovedisplayskip}{5pt}
\setlength{\belowdisplayskip}{5pt}
\begin{equation}
IRF_{\bullet j}(\ell) = (S_{n} \mathcal{C}_{\Pi}^\ell S_{n}^\prime) \, 
H_{\bullet 1} e_{j}, \quad 1 \leq j \leq k,  \label{IRF_j}
\end{equation}
}
where $S_{n} = (I_{n}, \, 0_{n \times n(p-1)}\,)$ is a selection matrix, and 
$e_{j}$ is a $k \times 1$ vector with a \textquotedblleft 1\textquotedblright\ 
in the $j$-th position and zeros elsewhere. Equation \eqref{IRF_j} represents the \textquotedblleft absolute\textquotedblright\ 
responses to one-standard-deviation target shocks.  In the special case where $k=1$ (a single target shock), it is convenient to refer to the relative on-impact responses, defined as $H_{2,1}^{\text{rel}} := H_{2,1}\,/\,h_{1,1}$, where $h_{1,1}$ is the (1,1) entry of $H$. 
$H_{2,1}^{\text{rel}}$ incorporates the unit effect normalization that ensures that the on-impact 
response of $Y_{1,t}$ to the impulse $\varepsilon_{1,t}$ is equal to 1, where $Y_{1,t}$ 
is 
the first variable in the VAR. Therefore, for $k=1$, the ``relative'' impulse responses under the unit effect normalization replace $H_{\bullet 1}$ in \eqref{IRF_j} with the column vector $H_{\bullet 1}/h_{1,1}$. 
While the reduced form parameters in the companion matrix $%
\mathcal{C}_{\Pi }$ can be easily estimated with ordinary least squares, 
the identification of the on-impact coefficients in $H_{\bullet
1} $ (or $H_{2,1}^{rel}$) is challenging in the absence of auxiliary information. 

The solution provided by the \textquotedblleft external
instruments approach\textquotedblright\ is to consider an $r\times 1$ vector
of variables external to the VAR, say $z_{t}$, $r\geq k,$ which satisfy the
following conditions:\vspace{-20pt}
{\setlength{\abovedisplayskip}{5pt}
\setlength{\belowdisplayskip}{5pt}
\begin{align}
\text{relevance:  } & \qquad \mathbb{E}(z_{t} \, \varepsilon _{1,t}^{\prime }) \,= \Phi\, , 
\qquad rank[\Phi ]=k \,,   \label{C1}\\ 
\text{exogeneity:  } & \qquad \mathbb{E}(z_{t}\, \varepsilon _{2,t}^{\prime }) \,= 0_{r\times (n-k)}\, , \label{C2}
\end{align} }
where $\Phi $ is an $r\times k$ matrix of relevance parameters. Combining Equation \eqref{partition_B} with conditions \eqref{C1}--\eqref{C2} yields the moment conditions: {\setlength{\abovedisplayskip}{5pt}
\setlength{\belowdisplayskip}{5pt}
\begin{equation}
\mathbb{E}(u_{t}z_{t}^{\prime })=\Sigma _{u,z}:= \left( 
\begin{array}{c}
\Sigma _{u_{1},z} \\ 
\Sigma _{u_{2},z}%
\end{array}%
\right) =H_{\bullet 1}\Phi ^{\prime } = \left( 
\begin{array}{c} 
H_{1,1}\Phi ^{\prime } \\ 
H_{2,1}\Phi ^{\prime }%
\end{array}%
\right) \text{ \ \ }%
\begin{array}{l}
k\times r \\ 
(n-k)\times r%
\end{array}
\label{cov_uz}
\end{equation}}
which represent the key ingredients of the proxy-SVAR approach; see 
\cite{MertensRavn2013} and \cite{StockWatson2018}.

For our purposes, define matrix $R_{z} := (\Phi,\,\Upsilon)$, where $\Upsilon$ is the $r$-by-$(n-k)$ matrix of contamination parameters, and set $\Upsilon=0_{r\times (n-k)}$. A convenient summary of the proxy conditions \eqref{C1}--\eqref{C2} 
is captured by the linear measurement error model:
{\setlength{\abovedisplayskip}{5pt}
\setlength{\belowdisplayskip}{5pt}
\begin{equation}\label{linear_measurement_error0}
z_t = R_{z} \varepsilon_t + \Omega_{tr} \, \zeta_t, 
\end{equation}}
where $\zeta_t$ is a normalized $r$-dimensional random variable with covariance matrix $\mathbb{E}(\zeta_t \zeta_t^\prime) = I_r$, and assumed to be uncorrelated with the structural shocks;  $\Omega_{tr}$ is a scaling matrix such that $\Omega = \Omega_{tr} \Omega_{tr}'$ can be interpreted as the $r\times r$ covariance matrix of the  measurement errors. Therefore, the covariance 
matrix of the proxies is $\Sigma_z := \mathbb{E}(z_t z_t^\prime) = R_{z} R_{z}^\prime+ 
\Omega$. The specification 
of $R_{z}$ in \eqref{linear_measurement_error0} is flexible and allows: $\Upsilon = 0_{r \times (n-k)}$ under instrument 
exogeneity (thus, leading to the moment conditions \eqref{cov_uz}), and $\Upsilon \neq 0_{r \times (n-k)}$ when instruments are contaminated 
(i.e., when Condition  \eqref{C2} does not hold).

Given \eqref{linear_measurement_error0}, the proxy-SVAR in \eqref{VAR-RF2}\ can be expressed as:
{\setlength{\abovedisplayskip}{5pt}
\setlength{\belowdisplayskip}{5pt}
\begin{align}
\underset{W_{t}}{\underbrace{\left( 
\begin{array}{c}
Y_{t} \\ 
z_{t}%
\end{array}%
\right) }} & =\underset{\Gamma }{\underbrace{\left( 
\begin{array}{c}
\Pi \\ 
0_{r\times np}%
\end{array}%
\right) }} \, X_{t}+\underset{\eta _{t}}{\underbrace{\left( 
\begin{array}{c}
u_{t} \\ 
z_{t}%
\end{array}%
\right) }}  \label{eqn:AC1}\\  \underset{\eta _{t}}{\underbrace{\left( 
\begin{array}{c}
u_{t} \\ 
z_{t}%
\end{array}%
\right) }}& =\underset{G}{\underbrace{\left( 
\begin{array}{cc}
H & 0_{n\times r}\\ 
R_{z} & \Omega _{tr }%
\end{array}%
\right) }}\underset{\xi _{t}}{\underbrace{\left( 
\;\begin{array}{c}
\varepsilon _{t} \\ 
\zeta _{t}%
\end{array}%
\right) }} = \left( 
\begin{array}{ccc}
H_{\bullet 1} & H_{\bullet 2} &  0_{n\times r}\\ 
\Phi & \Upsilon & \Omega _{tr }%
\end{array}%
\right)\; \left( 
\begin{array}{c}
\varepsilon _{1,t} \\ 
\varepsilon _{2,t} \\ 
\zeta _{t}%
\end{array}%
\right)  \label{eqn:AC2}
\end{align}}
where $R_{z}$ contains the relevance parameters and possibly the contamination parameters (when $\Upsilon\neq 0_{r\times(n-k)}$), and $\xi _{t}$ stacks the structural shocks $\varepsilon_t$ and
the normalized measurement errors $\zeta_t$, such that $\mathbb{E}(\xi _{t}\xi _{t}^{\prime })=I_{n+r}$;\ see, e.g., \cite{AngeliniFanelli2019},
\cite{Ariasetal2021} and \cite{GiacominiKitagawaRead2022} for similar representations.  In equation \eqref{linear_measurement_error0}, it is assumed that the proxies $z_{t}$
are expressed in their  \textit{innovations form},
meaning they are generated by a serially uncorrelated process. This
condition can be relaxed. The top-right zero block within matrix $G$ in Equation \eqref{eqn:AC2} reflects the fact that the instrument measurement errors have by construction no effect on the variables $Y_t$, i.e. $\mathbb{E}[u_t\zeta_t'] = 0_{n\times r}$ with probability one.
 
The vector $\eta _{t}$\ in (\ref{eqn:AC2})\ incorporates the VAR\
innovations and the proxies; the corresponding $(n+r)\times (n+r)$ covariance
matrix is $\Sigma _{\eta }:=\mathbb{E}(\eta _{t}\eta _{t}^{\prime
})=GG^{\prime }$. The matrix $G$ in \eqref{eqn:AC2} plays a key role for our
analysis. In its more general form, it gives rise to the set of covariance
restrictions: 
{\setlength{\abovedisplayskip}{5pt}
\setlength{\belowdisplayskip}{5pt}
\begin{equation}
\Sigma _{\eta } := \left( 
\begin{array}{cc}
\Sigma _{u} & \Sigma _{u,z} \\ 
\Sigma _{z,u} & \Sigma _{z}%
\end{array}%
\right) 
=\left( 
\begin{array}{cc}
H_{\bullet 1}H_{\bullet 1}^{\prime }+H_{\bullet 2}H_{\bullet 2}^{\prime } & 
H_{\bullet 1}\Phi ^{\prime }+H_{\bullet 2}\Upsilon ^{\prime } \\ 
\Phi H_{\bullet 1}^{\prime }+\Upsilon H_{\bullet 2}^{\prime } & \Phi \Phi
^{\prime }+\Upsilon \Upsilon ^{\prime }+\Omega%
\end{array}%
\right) ,  \label{structure-G}
\end{equation}}
which incorporate four cases of interest for the proxies $z_{t}$. Specifically, assuming a drifting DGP characterized by sequences of models where 
$\mathbb{E}\left(z_t \varepsilon_{1,t}^\prime\right) = \Phi_T.$ Instruments are defined: c.(i) strong and exogenous if $\Phi _{T}\rightarrow $ $\Phi $ and conditions \eqref{C1}--\eqref{C2} hold; c.(ii) local-to-zero and exogenous if $\Phi _{T}=T^{-1/2}C$, $C$ being an $r\times
k$ matrix with finite norm, $\left\Vert C\right\Vert <\infty $, and Condition \eqref{C2} holds; c.(iii) strong and contaminated if $\Phi _{T}\rightarrow $ $\Phi $, Condition \eqref{C1} holds, and $\Upsilon \neq 0_{r\times (n-k)};$ c.(iv) local-to-zero and contaminated if $\Phi _{T}=T^{-1/2}C$, $C$ being
an $r\times k$ matrix with finite norm, $\left\Vert C\right\Vert <\infty $,
and $\Upsilon \neq 0_{r\times (n-k)}.$

Starting from the premise that valid external instruments, as defined by \cite{StockWatson2018}, satisfy definition c.(i), we will henceforth consider the external instruments defined in definitions c.(ii)--c.(iv) as \textit{invalid}. Specifically, definition c.(iv) represents our broadest interpretation of invalid proxies, encompassing scenarios where the external instruments \( z_{t} \) exhibit weak correlations with the target shocks, as described by \cite{StaigerStock1997}, while also being correlated with some or all non-target shocks.

\subsection{DGP\ and assumptions}

\label{Section_DGP}In this section, we summarize our main assumptions. We
relax the hypothesis that the VAR\ parameters $(\Pi $, $\Sigma _{u})$ and,
possibly, the external instruments parameters $(R_{z},\Omega_{tr})$
are time-invariant over the sample $W_{1},...,W_{T}$. The two
assumptions that follow introduce a structural break in \eqref{eqn:AC1}--\eqref{eqn:AC2} and establish the regularity conditions under which our analysis applies. 

Hereafter, superscript ``(0)\textquotedblright\ denotes parameter vectors/matrices evaluated 
at their true (DGP) value. Notation $\mathbb{I}\left( \bullet \right)$ represents the indicator operator.

\begin{assumption}[Proxy-SVAR with a shift in volatility]
\label{ass:break} Let $T_{B}$ be a break date, with \mbox{$1<T_{B}<T$}. The reduced form
associated with the proxy-SVAR in \eqref{eqn:AC1}--\eqref{eqn:AC2} belongs to the DGP: \vspace{-10pt}
{\setlength{\abovedisplayskip}{-0pt}
\setlength{\belowdisplayskip}{0pt}
\begin{equation}
W_{t}=\Gamma (t)X_{t}+\eta _{t}, \qquad \Sigma _{\eta }(t):=\mathbb{E}%
(\eta _{t}\eta _{t}^{\prime })\text{ \ , \ }t=1,...,T \label{VAR_RF_with_one_break}
\end{equation}}\vspace{-35pt}
{\setlength{\belowdisplayskip}{-5pt}
{\setlength{\abovedisplayskip}{0pt}
\begin{align*} \text{ where } \qquad 
\Gamma (t) &  :=\; \Gamma _{1}\cdot \mathbb{I}\left( t\leq T_{B}\right)
\; +  \; \Gamma _{2}\cdot \mathbb{I}\left( t \geq T_{B}+1\right) \\
\Sigma _{\eta }(t)&:=  \; \Sigma _{\eta ,1}\cdot \mathbb{I}\left( t\leq
T_{B}\right) \;+ \; \Sigma _{\eta ,2}\cdot \mathbb{I}\left( t\geq T_{B}+1\right),
\end{align*}
}
}
and\newline
(i) the process $\{\eta _{t}\}$, $\eta
_{t}:=(u_{t}^{\prime },z_{t}^{\prime })^{\prime }$, is $\alpha $-mixing on
both samples $W_{1},...,W_{T_{B}}$ and $W_{T_{B}+1},...,W_{T}$, meaning that
it satisfies the conditions in Assumption 2.1 in \cite{BrueggemannJentschTrenkler2016}; furthermore, the process $\{\eta _{t}\}$ has absolutely
summable cumulants up to order eight on both samples $W_{1},...,W_{T_{B}}$
and $W_{T_{B}+1},...,W_{T}$;\newline
(ii) $\Sigma _{\eta ,1}<\infty $ and $\Sigma _{\eta
,2}<\infty $ are positive definite;\newline
(iii) each regime-dependent parameter $\big(\Gamma
_{i}^{(0)},\Sigma _{\eta ,i}^{(0)}\big)$, $i=1,2$, corresponds to a covariance
stationary VAR\ process for $W_{t}$;\newline
(iv) $\Sigma _{\eta ,2}^{(0)}\neq \Sigma _{\eta
,1}^{(0)}.$
\end{assumption}

\begin{assumption}[Break Date]
\label{ass:breakdate} $T_{B}=\lfloor \tau^{(0)} _{B}  T\rfloor$, with $\tau^{(0)}
_{B}   \in (0,1)$ being the fraction of observations in the first volatility regime.
\end{assumption}

Assumption \ref{ass:break} postulates that the\ unconditional error
covariance matrix $\Sigma _{\eta }$\ shifts at the break date $T_{B}$.
Despite this shift, the system remains \textit{stable} within the two volatility regimes in the following sense. First, the process that generates the VAR\ disturbances and the
proxies, $\left\{ \eta _{t}\right\} $, is $\alpha $-mixing and has
absolutely summable cumulants up to order eight (Assumption \ref{ass:break}.(i)) in both regimes. The $\alpha$-mixing condition for $\eta_t$ encompasses scenarios where, for example, the VAR disturbances and proxies are driven by conditionally heteroskedastic processes (such as GARCH) and/or the proxies are generated by zero-censored processes \citep{JentschLunsford2022}.
Assumption \ref{ass:break}.(i) is a technical requirement essential to
guarantee Moving Block Bootstrap (MBB) consistency 
\citep[see Assumption 2.4 in][]{JentschLunsford2022}. Second, the unconditional covariance matrices $%
\Sigma _{\eta ,1}$ and $\Sigma _{\eta ,2}$ are finite and positive definite (Assumption \ref{ass:break}.(ii)), and the VAR for $W_{t}$\ is asymptotically stable in both volatility
regimes (Assumption \ref{ass:break}.(iii)). Assumption \ref{ass:break}.(iv)
establishes that the unconditional covariance matrices $\Sigma _{\eta ,1}$
(pre-break period) and $\Sigma _{\eta ,2}$ (post-break period) are different. 
 Finally,  Assumption \ref{ass:break} is consistent with scenarios where the autoregressive parameters may change ($\Gamma_1 \neq \Gamma_2$) or remain constant ($\Gamma_1 = \Gamma_2$) across volatility regimes \citep[see e.g.][]{bacchiocchi2024svarsbreaksidentificationinference}.

Assumption \ref{ass:breakdate} posits that the number of observations in
each regime increases as the sample size increases allowing for the asymptotic theory developed in, e.g.,  
\cite{Bai2000}. Under  an additional MDS condition for $\xi_t$, such that $\mathbb{E}(\xi _{t}\mid \mathcal{I}_{t-1})=0_{n+r},$ together with $\mathbb{E}(\xi_t\,\xi_t'|\mathcal{I}_{t-1}) = I_{n+r}$ and $\sup_t\mathbb{E}(||\xi_t||^{4+\epsilon})<\infty$ for $\epsilon>0$, Assumptions \ref{ass:break}--\ref{ass:breakdate} ensure that $T(\hat{\tau}_{B}-\tau _{B}^{(0)})=O_{\mathbb{P}}(1)$, where $\hat{\tau}_{B}$ is the change-point
estimator discussed, e.g., in \cite{Bai2000}.
This implies that $\tau_B$ can be consistently estimated from the data and converges at a rate faster than $\sqrt{T}$, which is the convergence rate of the estimator of the parameters $(\Pi, \Sigma_\eta)$. Consequently, there are no concerns about pre-testing bias when constructing confidence intervals for the target IRFs. Interestingly, in many macroeconomic contexts, distinct volatility regimes are often readily observable. These regimes are frequently associated with economic crises or significant policy changes. Importantly, whether this break date is estimated from the data or assumed to be known, the underlying cause of the volatility shift does not need to be identified. For example, in Section \ref{Section_empirical_illustration}, we incorporate the volatility reduction associated with the Great Moderation into a fiscal proxy-SVAR, even though the exact cause of this moderation --whether due to ``good policy'' or ``good luck''-- remains debated and is likely unrelated to fiscal policy actions. 

For tractability, we restrict attention to a single, permanent shift in volatility; Section S.4 in the supplementary material generalizes the framework to multiple breaks. This shift delineates non-recurrent volatility regimes that correspond to distinct macroeconomic regimes. For instance, a passive monetary policy phase (regime 1) may be followed by an active monetary policy phase (regime 2) and later by a zero-lower-bound phase (regime 3), without assuming the system reverts to previous states. This flexibility requires a more sophisticated set of identifying restrictions and a larger number of parameters to estimate, as detailed in the sections that follow.

\subsection{Estimation of Target IRFs Ignoring Volatility Shifts}
\label{Section_Estimation_ignoring_break}
The main implication of Assumption \ref{ass:break} is that the subsets of observations $(W_{1}, \dots, W_{T_{B}})$ and $(W_{T_{B}+1}, \dots, W_{T})$ are characterized by two distinct error covariance matrices, $\Sigma_{\eta,1}$ and $\Sigma_{\eta,2}$, respectively. In this section, we investigate whether and under what conditions the target IRFs can be estimated  consistently using the instruments $z_t$ alone, despite the volatility shift, without the need to split the sample. We also examine the scenarios where the proxy-SVAR estimated over the whole sample is not consistent.

For simplicity, we posit that under Assumptions \ref{ass:break}--\ref{ass:breakdate}, the parameters in $(R_z, \Omega_{tr})$ remain constant across both volatility regimes. Specifically, we assume that there is no structural break in the process generating the proxies, which are generated from Equation \eqref{linear_measurement_error0}.
Thus, the structural break exclusively impacts the VAR error covariance matrix. This assumption implicitly imposes stability restrictions on the DGP
for the instruments as, e.g., in \cite{antoine2018efficient} and \cite{antoine2024efficient}, who estimate IV regressions with change-points. In this
context, the condition $\Sigma _{\eta ,2}\neq \Sigma _{\eta ,1}$ can be
solely ascribed to the change in the VAR error covariance matrix, $\Sigma
_{u,2}\neq \Sigma _{u,1}$. The hypothesis of no shifts in the parameters $%
(R_{z},\Omega_{tr}$) will be relaxed in  Section \ref%
{Section_stability_restrictions_approach}.

To streamline the presentation, we introduce the following notation. For any matrix $A$, let $\Delta_A$ denote a matrix of the same dimensions as $A$, whose nonzero elements, under Assumptions \ref{ass:break}--\ref{ass:breakdate}, capture the changes in $A$ from the first to the second regime. Formally, we define $\Delta_A := A^{(2)} - A^{(1)},$
where $A^{(1)} = A$ represents the matrix \(A\) in the first regime, and \(A^{(2)} = A + \Delta_A\) denotes the corresponding matrix in the second volatility regime. Then, the volatility shiFt $\Sigma _{u,2}\neq \Sigma _{u,1}$ is modeled by the $n(n+1)$ moment conditions: 
{\setlength{\abovedisplayskip}{5pt}
\setlength{\belowdisplayskip}{5pt}
\begin{equation}
\begin{array}{ll}
\Sigma _{u,1}=HH^{\prime }, & t\leq T_{B}, \\ 
\Sigma _{u,2}=(H+\Delta _{H}) \, \Lambda \, (H+\Delta _{H})^{\prime }, & t\geq
T_{B}+1,%
\end{array}
\label{BF-approach}
\end{equation}}%
with $H = H^{(1)}$, and $H^{(2)} = H + \Delta_H$, where $\Delta_H$ is an $n \times n$ matrix whose nonzero elements capture potential variations in the on-impact coefficients of the matrix $H$ during the shift from the first to the second volatility regime. Additionally, the matrix $\Lambda \equiv dg(\Lambda)$ is an $n$-dimensional diagonal matrix with positive entries on the diagonal. The nonzero elements of $\Lambda$ are typically interpreted as the relative changes in the variances of the structural shocks in the second volatility regime compared to the first, where the shocks are normalized to have unit variance.
A diagonal element of $\Lambda $
equal to 1 indicates that the variance of the corresponding shock in $\varepsilon_{t}$ remains constant across the two volatility regimes.

Moment conditions \eqref{BF-approach} imply that the  volatility shift depends on two components: (i) 
the nonzero coefficients in the matrix $\Delta _{H}$; (ii) the  
diagonal entries of matrix $\Lambda$ that are different from 1. According to \eqref{BF-approach}, when $\Delta_{H}=0_{n\times n}$, the change in volatility solely depends on the shift of variance of the structural shocks, and the dynamic causal effects remain unchanged across the two volatility regimes (only the scale of the response varies); see, e.g. \cite{LanneLutkepohl2008}. 
Conversely, when $\Delta _{H}\neq 0_{n\times n}$, equation \eqref{BF-approach} implies a scenario where the
volatility break alters the responses of the variables to the shocks other than the relative variances of the latter; see, e.g., \cite{BacchiocchiFanelli2015}, \cite{angelinietal2019}. 

Under Assumptions \ref{ass:break}--\ref{ass:breakdate}, the proxy-SVAR features the following  $\ell $-period ahead (absolute) responses
of $Y_{t+\ell }$ to one-standard deviation $j$-th target shock in $\varepsilon _{1,t}$, $1\leq j\leq k$: 
{\setlength{\abovedisplayskip}{5pt}
\setlength{\belowdisplayskip}{5pt}
\begin{equation}
IRF_{\bullet j}(t,\ell )=\left\{ 
\begin{array}{ll}
(S_{n}(\mathcal{C}_{\Pi })^{\ell }S_{n}^{\prime })H_{\bullet 1}e_{j}, & t\leq
T_{B}, \\ 
(S_{n}(\mathcal{C}_{\Pi })^{\ell }S_{n}^{\prime })(H_{\bullet 1}+\Delta
_{H_{\bullet 1}})\left( \Lambda _{\bullet 1}^{1/2}\right) e_{j}, & t\geq
T_{B}+1,%
\end{array}%
\right.   \label{abs_irfs_bf}
\end{equation}%
}
where matrices $\Delta_{H_{\bullet 1}}$ and $\Lambda_{\bullet 1}$ denote the corresponding $n \times k$ and $k \times k$ top-left blocks of $\Delta_H$ and $\Lambda$, respectively. Equation \eqref{abs_irfs_bf} assumes that the VAR slope parameters, $\mathcal{C}_{\Pi}$, remain constant across both volatility regimes. This assumption is consistent with Assumptions \ref{ass:break} and \ref{ass:breakdate}. However, allowing for regime-dependent slope parameters is straightforward: replace $\mathcal{C}_{\Pi}$ in \eqref{abs_irfs_bf} with $\mathcal{C}_{\Pi,1}$ for $t \leq T_B$ and $\mathcal{C}_{\Pi,2}$ for $t \geq T_B + 1$. For $k=1$ (single target shock),  the relative, normalized target IRFs are:
{\setlength{\abovedisplayskip}{5pt}
\setlength{\belowdisplayskip}{5pt}\begin{equation}
\frac{IRF_{\bullet 1}(t,\ell )}{IRF_{1,1}(t,0)}=\left\{ 
\begin{array}{lc}
(S_{n}(\mathcal{C}_{\Pi })^{\ell }S_{n}^{\prime })\left( 
\begin{array}{c}
1 \\ 
H_{2,1}^{rel}%
\end{array}%
\right), & t\leq T_{B}, \\ 
(S_{n}(\mathcal{C}_{\Pi })^{\ell }S_{n}^{\prime })\left( 
\begin{array}{c}
1 \\ 
\frac{H_{2,1}+\Delta _{H_{2,1}}}{h_{1,1}+\Delta _{h_{1,1}}}%
\end{array}%
\right), & t\geq T_{B}+1,%
\end{array}%
\right.  \label{rel_irf_bf}
\end{equation}
}%
where $\Delta _{H_{\bullet 1}}$ has been partitioned as $\Delta _{H_{\bullet
1}}=(\Delta _{h_{1,1}},\Delta _{H_{2,1}}^{\prime })^{\prime }$. Scalars $h_{1,1}$ and $\Delta _{h_{1,1}}$ correspond to the (1,1) entries of $H$ and $\Delta_H$, respectively.  Relative (normalized) target IRFs can be also generalized to the case  $k>1$, situation we address in the empirical illustration presented in Section \ref{Section_empirical_illustration}.  The key fact about normalized IRFs, compared to absolute IRFs,  is that relative responses do not involve the parameters in $\Lambda_{\bullet 1}$, i.e. the possible changes in the variances of the structural shocks. This implies that when practitioners are interested in relative responses, identifying restrictions on $\Lambda_{\bullet 1}$ are unnecessary.

With all the necessary components in place, we can now establish our main results regarding the estimation of the target IRFs in equations~\eqref{abs_irfs_bf} and~\eqref{rel_irf_bf} using only external instruments, while ignoring the volatility shift in the DGP. 
For our purposes, it is sufficient to study the large sample behavior of the estimator $\hat{\Sigma}_{u,z}= \frac{1}{T}\sum_{t=1}^{T}\hat{u}_{t}z_{t}^{\prime }.$ Henceforth, \textquotedblleft $\overset{\mathbb{P}}{\rightarrow }$\textquotedblright\
denotes convergences in probability.  

\begin{proposition}[Convergence of $\hat{\Sigma}_{u,z}$ in the presence of a volatility shift]
\label{Proposition 1} 
Under Assumptions \ref{ass:break} and \ref{ass:breakdate}, consider the proxy-SVAR model with $\Gamma(t) = \Gamma$  in Equation \eqref{VAR_RF_with_one_break} for all $t$, where the instrument parameters $(R_z, \Omega_{tr})$ remain constant across the two volatility regimes, as generated from Expression \eqref{linear_measurement_error0}. Further assume that $ \mathbb{E}\left(z_t \varepsilon_{1,t}'\right) = \Phi_T$,
where the proxies \( z_t \) satisfy the conditions c.(i), i.e. they are strong and exogenous. 

Then, under the volatility shifts featured by \eqref{BF-approach}, the
estimator $\hat{\Sigma}_{u,z} = \frac{1}{T}\sum_{t=1}^{T}\hat{u}_{t}z_{t}^{\prime }$ is such that:\newline
(i)  $\hat{\Sigma}_{u,z}\overset{\mathbb{P}}{\rightarrow } \Sigma_{u,z}^{(0)},$
with  
{\setlength{\abovedisplayskip}{5pt}
\setlength{\belowdisplayskip}{5pt}
\begin{equation*}
\Sigma_{u,z}^{(0)} \, =  \, \left[\tau _{B}^{(0)} \, H_{\bullet 1}^{(0)}\,\, +\, \left(1-\tau _{B}^{(0)}\right)\left(H_{\bullet 1}^{(0)} + \Delta^{(0)} _{H_{\bullet 1}}\right) \left( \Lambda _{\bullet 1}^{(0)}\right)^{1/2}\,\right]  \left(\Phi ^{(0)}\right)^{\prime }  ;
\end{equation*}%
}
(ii) $\hat{\Sigma}_{u_{2},z}\hat{\Sigma}_{u_{1},z}^{-1} \overset{\mathbb{P}}{\rightarrow } \Sigma_{u_{2},z}^{(0)}\left(\Sigma_{u_{1},z}^{(0)}\right)^{-1}\,,$ and for $k=1$,  
{\setlength{\abovedisplayskip}{5pt}
\setlength{\belowdisplayskip}{5pt}
\begin{equation*}
\Sigma_{u_{2},z}^{(0)} \big/ \Sigma_{u_{1},z}^{(0)} = 
\tau _{B}^{(0)}\,
H_{2,1}^{rel,{(0)}} \, +\, \left(1-\tau _{B}^{(0)}\right)\,
\frac{H_{2,1}^{(0)}+\Delta^{(0)}_{H_{2,1}}}{ h_{1,1}^{(0)} +\Delta^{(0)}_{h_{1,1}}} .
\end{equation*}
}
\end{proposition}

Proposition \ref{Proposition 1} establishes that ignoring the volatility shift the proxy-SVAR will estimate a convex combination  of the on-impact coefficients across the two regimes, with (positive) weights determined by the proportion of observations in each regime. Then, it is impossible, without further restrictions, to recover the on-impact parameters in $H_{\bullet 1}$ and $\Delta_{H_{\bullet 1}}$. Consequently, relying solely on external instruments generally fails to consistently estimate the IRFs both in \eqref{abs_irfs_bf} and \eqref{rel_irf_bf}. 

Section S.2 in the supplementary material specializes the results in Proposition \ref{Proposition 1} to the scenario where IRFs do not change across volatility regimes (a common assumption in the identification-through-heteroskedasticity approach).  Overall, the main takeaway from Proposition \ref{Proposition 1} is that not incorporating volatility shifts into proxy-SVARs may crucially invalidate inference. This is true, in particular, when it is difficult to justify the assumption that the target IRFs remain constant across macroeconomic regimes.

\subsection{Proxy-SVARs with Stability Restrictions}

\label{Section_stability_restrictions_approach}In this section, we introduce
our stability restrictions approach to the identification and estimation of
proxy-SVARs under a permanent, exogenous change in unconditional volatility.
Our focus is the point identification and estimation of the target IRFs as
defined in (\ref{abs_irfs_bf}). The reference proxy-SVAR is specified under 
Assumptions \ref{ass:break}-\ref{ass:breakdate}. Hereafter, shifts in $R_z,$ and $\Omega_{tr}$ are also considered. We first explore identification issues, then proceed discussing estimation and inference.

\subsubsection{Identification}

We consider an extension of moment conditions \eqref{BF-approach} to the covariances of $\eta_t$, the vector collecting VAR\ innovations and proxies. Thus, under Assumptions \ref{ass:break}--\ref{ass:breakdate}, Equation \eqref{eqn:AC2} reads
{\setlength{\abovedisplayskip}{5pt}
\setlength{\belowdisplayskip}{5pt}
\begin{equation}
\eta _{t}= \, G\, \xi _{t}\cdot \mathbb{I}\left( t\leq T_{B}\right)\, +\, (G+\Delta
_{G})\left( \Psi ^{1/2}\right) \, \xi _{t}\cdot \mathbb{I}\left( t\geq T_{B}+1\right)
,  \label{model_changes}
\end{equation}}
where  
$\Psi\equiv dg(\Psi ) = diag(\Lambda,\, \Lambda_\zeta)$ is a block diagonal matrix with distinct, positive elements on the diagonal that reflect the changes in the relative variances of the elements in $\xi _{t}$ from the first to the second volatility regime, with $\Lambda_\zeta$ representing the relative volatility shift of measurement errors. Similarly to \eqref{BF-approach}, the implied moment conditions are:%
{\setlength{\abovedisplayskip}{5pt}
\setlength{\belowdisplayskip}{5pt}
\begin{equation}
\begin{array}{ll}
\Sigma _{\eta ,1}\,=\,G\, G^{\prime }, &t\leq T_{B}, \\ 
\Sigma _{\eta ,2}\,=\,(G+\Delta _{G})\, \Psi \, (G+\Delta _{G})^{\prime }, &%
t\geq T_{B}+1\, .%
\end{array}%
   \label{Moment_conditions_one_break_large}
\end{equation}
}
The change in the covariance matrix from $\Sigma_{\eta,1}$ to $\Sigma_{\eta,2}$ can be ascribed to two components: (i) changes in the
impact of the shocks on the variables and instruments'  relevance and contamination, captured by the nonzero elements in\ $\Delta _{G}$; (ii) changes in the relative
variance of the structural shocks and instruments' measurement error, captured by the diagonal elements of $\Psi $. In (\ref{Moment_conditions_one_break_large}), also instrument relevance,
exogeneity and variance of $\zeta_t$ can potentially
shift. In its general form, the structure of the matrix $G+\Delta _{G}$ in
the second volatility regime is given by%
\begin{equation}
G+\Delta _{G} 
 = \left( 
\begin{array}{ccc}
H_{\bullet 1}+\Delta _{H_{\bullet 1}} & H_{\bullet 2}+\Delta _{H_{\bullet 2}}
& 0_{n\times r} \\ 
\Phi +\Delta _{\Phi } & \Upsilon +\Delta _{\Upsilon } & \Omega _{tr}+\Delta _{\Omega _{tr }}%
\end{array}%
\right),  \label{eqn:structure_G+THETA}
\end{equation}%
so that it is seen that while the nonzero elements in $\Delta _{H}$ account
for possible changes in the on-impact coefficients, the nonzero elements in $%
\Delta _{R_{z }}:=(\Delta _{\Phi }$, $\Delta _{\Upsilon })$ and $\Delta
_{\Omega _{tr}}$\ reflect variations in the parameters governing proxy
properties, namely changes in relevance, contamination and measurement
errors' variability. The zero restrictions within matrices $G$ and $\Delta _{G}$ in (\ref{eqn:structure_G+THETA})\ and the diagonal structure of $\Psi $ do not
necessarily guarantee that the moment conditions in (\ref%
{Moment_conditions_one_break_large}) identify the proxy-SVAR. In principle,
the reduced form parameters in $\Sigma _{\eta ,1}$ and $\Sigma _{\eta ,2}$
might be fewer than the nonzero elements in $G$, $\Delta _{G}$ and $\Psi $.
Alternatively, the order condition might hold but not the rank condition for
identification. As in \cite{MagnussonMavroeidis2014}, 
point-identification can be achieved 
by imposing a set of (linear)
constraints on $G$, $\Delta _{G}$ and $\Psi $ that we express in explicit
form: {\setlength{\abovedisplayskip}{5pt}
\setlength{\belowdisplayskip}{5pt}
\begin{eqnarray}\label{eqn:R1}
vec(G) = S_{G} \, \gamma +s_{G} \, ,   \quad  \quad 
vec(\Delta _{G}) =S_{\Delta _{G}}\, \delta +s_{\Delta _{G}} \, ,   \, \quad \quad 
vecd(\Psi)  =  S_{\Psi }\, \psi +s_{\Psi }\, .  
\end{eqnarray}%
}

In Equation \eqref{eqn:R1}, $S_G$ is a full column-rank $(n+r)^2 \times a$ selection matrix with $a \leq (n+r)^2$, mapping the $a$ unconstrained parameters in $G$ into the vector $\gamma$. Similarly, $S_{\Delta_G}$ ($ (n+r)^2 \times b$) selects the $b$ free parameters in $\Delta_G$, forming the vector $\delta$. 
Vectors $s_G$ and $s_{\Delta_G}$ ($ (n+r)^2 \times 1$) contain known elements of $G$ and $\Delta_G$.  $vecd(\bullet)$ is the $vec$ operator for diagonal matrices (see Section S.1 in the supplementary material), and $S_{\Psi }$ ($ (n+r) \times c$) selects the $c \leq (n+r)$ non-calibrated diagonal elements of $\Psi$, forming the vector $\psi$. Finally, $s_{\Psi}$ ($ (n+r) \times 1$) contains known elements of $\Psi$. The flexibility of the moment conditions in (\ref{Moment_conditions_one_break_large}) under the stability restrictions  (\ref{eqn:R1}) is illustrated by an example in Section S.3 of the supplementary material. Notably, \eqref{eqn:R1} does not require assumptions on non-target shocks. However, as shown in the empirical illustration, credible restrictions on their impact can be easily incorporated when available. It is important to note that restrictions in (\ref{eqn:R1})  envision a scenario in which the investigator possesses sound theoretical or empirical reasons to specify, e.g., which parameters  in the matrix $\Delta_G$ remain constant across volatility regimes. The example sketched in Section S.3 discusses an exactly identified model which provides general guidance for practitioners on how to specify the restrictions in \eqref{eqn:R1} when a priori information about stability restrictions is scant. 

Under the stability restrictions in \eqref{eqn:R1}, the moment conditions (\ref{Moment_conditions_one_break_large}) feature $(n+r)(n+r+1)$ reduced form
coefficients, $\sigma _{\eta ,1}=vech(\Sigma _{\eta ,1})$ and $\sigma _{\eta
,2}=vech(\Sigma _{\eta ,2})$, and $a+b+c$ free parameters in $\varsigma
=(\gamma ^{\prime },\delta ^{\prime },\psi' )^{\prime }$, respectively.  We
summarize the moment conditions (\ref{Moment_conditions_one_break_large}) by
the distance function: 
{\setlength{\abovedisplayskip}{5pt}
\setlength{\belowdisplayskip}{5pt}\begin{equation}
m(\sigma _{\eta },\varsigma )=\left( 
\begin{array}{l}
\sigma _{\eta ,1}-vech\big(G\, G^{\prime }\big) \\ 
\sigma _{\eta ,2}-vech\big((G+\Delta _{G})\, \Psi \, (G+\Delta _{G})^{\prime }\big)%
\end{array}%
\right)   \label{eqn:distance_function}
\end{equation}%
}
which establishes a mapping between the reduced form covariance parameters $%
\sigma _{\eta }:=(\sigma _{\eta ,1}^{\prime },\sigma _{\eta ,2}^{\prime
})^{\prime }$ and $\varsigma $.  Let $\theta$ (a sub-vector of $\varsigma$) denote the vector of parameters associated with the target IRFs in \eqref{abs_irfs_bf}. The elements of $\theta $ are specific components
of the vectors $\gamma $, $\delta $ and $\psi $, respectively. Finally, $\dim \theta\leq a+b+c.$  

The next proposition establishes the
necessary and sufficient conditions for the identification of $\varsigma $.
If $\varsigma $ is identified, the parameters of interest in $\theta $ are also
identified. In the following,  the matrix $\mathcal{F}_{\bullet }:=\frac{\partial vec(\bullet )%
}{\partial vecd(\bullet )^{\prime }}$ is defined in Section S.1 of the supplementary material.

\begin{proposition}[Identification under Stability Restrictions]
\label{prop:nec_suff_rank_cond_regime_dependent_irfs} Given the proxy-SVAR
specified under Assumptions \ref{ass:break}-\ref{ass:breakdate}, consider the moment
conditions in \eqref{eqn:distance_function} where $G$, $\Delta _{G}$ and $%
\Psi $ are restricted as in \eqref{eqn:R1}. Assume $\varsigma
_{0} \in \mathcal{P}_{\varsigma }$ is a regular point of the Jacobian matrix $\mathcal{J}(\varsigma):= {\partial m(\sigma _{\eta },\varsigma )}\big/{\partial \varsigma ^{\prime }}$.
Then, irrespective of instrument properties: \newline 
(i) a necessary and sufficient condition for the
(local) identification of $\varsigma_0$  is that $rank[\mathcal{J}(\varsigma)]=a+b+c$  in a neighborhood of $\varsigma_0$, where $\mathcal{J}%
(\varsigma)$ is $(n+r)(n+r+1)\times (a+b+c)$, defined by: 
{\setlength{\abovedisplayskip}{5pt}
\setlength{\belowdisplayskip}{5pt}
\begin{equation}
\mathcal{J}(\varsigma) = 
2\left( I_{2}\otimes D_{n+r}^{+}\right) 
 \begin{pmatrix}
\mathcal{J}_{1,\gamma } & \mathcal{J}_{1,\delta } & \mathcal{J}_{1,\psi } \\ 
\mathcal{J}_{2,\gamma } & \mathcal{J}_{2,\delta } & \mathcal{J}_{2,\psi }
\end{pmatrix} \,  diag\Bigg(S_{G},\; S_{\Delta _{G}}, \;\frac{1}{2}\mathcal{F}_{\Psi }S_{\Psi }  \Bigg)\, ,%
\label{NECESSARY_SUFFICIENT_MATRIX_b2}
\end{equation}
}
with 
{\setlength{\abovedisplayskip}{5pt}
\setlength{\belowdisplayskip}{5pt}
\begin{eqnarray*}
 \mathcal{J}_{1,\gamma } &=&G\otimes I_{n+r} \, , \quad   \mathcal{J}_{1,\delta
}=\mathcal{J}_{1,\psi }=0_{(n+r)^2\times(n+r)^2}\, ; \quad \mathcal{J}_{2,\gamma }=(G+\Delta _{G})\, \Psi \otimes I_{n+r}\, ; \\  
\mathcal{J}_{2,\delta }&=& (G+\Delta _{G})\, \Psi \otimes I_{n+r}\, ;  \qquad 
\mathcal{J}_{2,\psi }=  (G+\Delta _{G})\otimes (G+\Delta
_{G}) \;;
\end{eqnarray*}}
(ii) a necessary order condition is: 
{\setlength{\abovedisplayskip}{5pt}
\setlength{\belowdisplayskip}{5pt}
\begin{equation}
(a+b+c)\leq (n+r)(n+r+1).  \label{NECESSARY_CONDITION_b}
\end{equation}} 
(iii) If $\Delta _{G}=0_{(n+r)\times (n+r)}$, $\varsigma :=(\gamma ^{\prime },\, \psi' )^{\prime }$ (set $\delta
=0_{b\times 1}$), the Jacobian collapses to%
{\setlength{\abovedisplayskip}{5pt}
\setlength{\belowdisplayskip}{5pt}
\begin{equation}
\mathcal{J}(\varsigma )=2(I_{2}\otimes D_{n+r}^{+})\left( 
\begin{array}{cc}
G\otimes I_{n+r} & 0_{(n+r)^2\times(n+r)^2} \\ 
G\, \Psi \otimes I_{n+r} & G\otimes G%
\end{array}%
\right) \left( 
\begin{array}{cc}
S_{G} &  0_{(n+r)^2\times c} \\ 
0_{(n+r)^2\times a}  & \frac{1}{2}\mathcal{F}_{\Psi} S_{\Psi}%
\end{array}%
\right) \; , \label{NECESSARY_CONDITION_c}
\end{equation}%
}
and a necessary and sufficient rank condition for the (local) identification
of $\varsigma_0$ is that $rank[\mathcal{J}%
(\varsigma)]=a+c$ in a neighborhood of $\varsigma_0$.
\end{proposition}

The interesting feature of Proposition \ref%
{prop:nec_suff_rank_cond_regime_dependent_irfs} is that the rank condition
holds regardless of instrument properties. This implies that if the volatility shift is sufficiently
informative and stability restrictions correctly specified, possible
breakdowns of proxy relevance and exogeneity do not invalidate the
identifiability of the model. Intuitively, instrument relevance is not strictly
necessary for identification because, whatever the properties of the
limiting matrix $\Phi $, the rank of the Jacobian $\mathcal{J}(\varsigma )$
in \eqref{NECESSARY_SUFFICIENT_MATRIX_b2} remains unaffected under sequences 
$\Phi _{T}\rightarrow $ $\Phi $. This implies that even in cases of invalid
proxies that satisfy the conditions c.(ii)-c.(iv), the
proxy-SVAR is still identifiable and asymptotic inference is standard. 
On the other hand, the exogeneity condition \eqref{C2} can be relaxed because the moment conditions implied by the shifts in volatility are informative also on the non-target shocks other than the
target shocks. This means that also the parameters in $H_{\bullet 2}$ and in 
$\Delta _{H_{\bullet 2}}$ are identified. Accordingly, if the
necessary and sufficient rank condition in Equation \eqref{NECESSARY_SUFFICIENT_MATRIX_b2}  
holds, the target structural shocks
can be recovered and estimated consistently even when the instruments are correlated with some non-target shocks. 
Therefore, the suggested approach remains valid even when the proxy-SVAR’s information set is properly expanded -for instance, as detailed in \cite{MertensRavn2014} to account for fiscal foresight phenomena- regardless of whether instruments possibly reflect information on anticipated shocks not explicitly modeled.

Finally, Proposition \ref{prop:nec_suff_rank_cond_regime_dependent_irfs}.(iii) clearly demonstrates that the case with constant IRFs across
volatility regimes represents a special case of the more general framework addressed here. Ideally,
identification in Proposition \ref{prop:nec_suff_rank_cond_regime_dependent_irfs}.(iii) can also be achieved
by leaving the matrix $G$ completely unrestricted, corresponding to the case where $S_{G}=I_{(n+r)^2}$ in \eqref{NECESSARY_CONDITION_c}. However, this would lead to efficiency losses, as the top-right block in $G$  is inherently restricted to zero because of the measurement error term (see Equation \eqref{eqn:AC2}). Proposition \ref{prop:nec_suff_rank_cond_regime_dependent_irfs}.(iii) helps to rationalize many
results presented in \cite{SchlaakRiethPodstawski2023} through Monte Carlo simulations.

In our framework, identification can fail when the volatility shift is too small to offset possible instrument invalidity; i.e., when the difference between the two regime‑specific covariance matrices, $\Sigma _{\eta ,2}-\Sigma _{\eta ,1}$, is negligible. This situation arises if Assumption~\ref{ass:break}.(iv) is violated: e.g. in large samples the sup‑norm distance $||\Sigma _{\eta ,2}-\Sigma _{\eta ,1}||_{\infty}$
 tends to zero, producing a weak volatility shift in the sense of \citet{Lewis2022}. We illustrate the consequences of this weak‑break scenario in Section S.5 of the supplementary material.

\subsubsection{Estimation}\label{sec:estimation}

Under the conditions of Proposition \ref%
{prop:nec_suff_rank_cond_regime_dependent_irfs}, the parameters $\varsigma $%
 , hence $\theta $, can be estimated by CMD. 
Under Assumptions \ref{ass:break}-\ref{ass:breakdate}, it holds the
asymptotic normality result: 
{\setlength{\abovedisplayskip}{5pt}
\setlength{\belowdisplayskip}{5pt}
\begin{eqnarray}
 \sqrt{T}\Big(\hat{\sigma}_{\eta }-\sigma _{\eta}^{(0)}\Big)  \overset{d}{\rightarrow }N\Big(0,\, V_{\sigma _{\eta }}\Big) 
 \label{asymptotic_normality_sigmas},  \qquad \qquad   \text{with }\; V_{\sigma _{\eta }} := diag\Big(V_{\sigma _{\eta ,1}},\; V_{\sigma _{\eta ,2}}\Big), 
 \nonumber 
\end{eqnarray}%
}
where $\sigma _{\eta}^{(0)}:=\big(\sigma _{\eta, 1}^{(0) \prime },\sigma _{\eta
,2}^{(0)\prime }\big)^{\prime }$ is the true value of $\sigma _{\eta }$. The
structure of the asymptotic covariance matrices $V_{\sigma _{\eta ,i}}$, $i=1,2$ is discussed in detail in \cite{BrueggemannJentschTrenkler2016} and references therein. We denote by $\hat{V}_{\sigma
_{\eta }}$ any consistent estimator of $V_{\sigma _{\eta }}$. Our candidate choice for $\hat{V}_{\sigma _{\eta }}$ is the MBB estimator which is consistent under Assumptions \ref{ass:break}--\ref{ass:breakdate} if bootstrap resampling is carried out within each volatility regime. 
Then, given $\hat{\sigma}_{\eta }$, a CMD estimator of $\varsigma $ follows from the minimization problem:
{\setlength{\abovedisplayskip}{5pt}
\setlength{\belowdisplayskip}{5pt}
\begin{equation}
\hat{\varsigma}_{T}:=\arg \min_{\varsigma \in \mathcal{P}_{\varsigma }}\, m_{T}(\hat{\sigma}_{\eta },\varsigma )^{\prime }\, \hat{V}_{\sigma _{\eta
}}^{-1}\, m_{T}(\hat{\sigma}_{\eta },\varsigma )  \label{CMD_key}
\end{equation}%
}
where $m_{T}(\hat{\sigma}_{\eta },\varsigma )^{\prime }:=\big(m_{T,1}(\hat{\sigma%
}_{\eta ,1},\varsigma )^{\prime },m_{T,2}(\hat{\sigma}_{\eta ,2},\varsigma
)^{\prime }\big)$ is the distance function defined in %
\eqref{eqn:distance_function} with $\sigma _{\eta }$ replaced by $\hat{%
\sigma}_{\eta }.$ 
The next proposition establishes asymptotic
properties of the CMD estimator of $\varsigma$ and $\theta.$

\begin{proposition}[Asymptotic properties of CMD estimator]
\label{prop:asym_dist} Let $\hat{\varsigma}_{T}$ and $\hat{\theta}_T$ be the CMD estimator of the
parameters $\varsigma $ obtained from (\ref{CMD_key}), and the corresponding
subvector of $\hat{\varsigma}_{T}$, respectively. Let $\varsigma^{(0)}$ be an interior of $%
\mathcal{P}_{\varsigma }$ (assumed compact), with $\theta^{(0)}\in \mathcal{P}%
_{\theta }\subseteq \mathcal{P}_{\varsigma }$. Under the conditions of
Proposition \ref{prop:nec_suff_rank_cond_regime_dependent_irfs}\, , \,  $\hat{\varsigma}_{T} \overset{\mathbb{P}}{\rightarrow} \varsigma^{(0)}$, $\hat{\theta}_{T} \overset{\mathbb{P}}{\rightarrow } \theta^{(0)},$ and
{\setlength{\abovedisplayskip}{5pt}
\setlength{\belowdisplayskip}{5pt}
\begin{eqnarray*}
\sqrt{T}\Big(\hat{\varsigma}_{T}-\varsigma^{(0)}\Big)\overset{d}{\rightarrow }%
N\Big(0 ,\, V_{\varsigma }\Big)\, ,  \qquad \qquad    \sqrt{T}\Big(\hat{\theta}_{T}-\theta^{(0)}\Big)\overset{d}{\rightarrow }N \Big(0 ,\, V_{\theta } \Big)\, ,
\end{eqnarray*}%
}
where $V_{\varsigma }:=\left( \mathcal{J}(\varsigma^{(0)})^{\prime }V_{\sigma
_{\eta }}^{-1}\mathcal{J}(\varsigma^{(0)})\right) ^{-1}$ and $V_{\theta }$ is
the corresponding block of $V_{\varsigma }$.
\end{proposition}

Proposition \ref{prop:asym_dist} establishes that under the stated
identification conditions, the target IRFs can be estimated consistently, and standard asymptotic inference holds regardless of proxy properties. The proposition also ensures that when there are more moment conditions than parameters, $(n+r)(n+r+1)>(a+b+c)$,  the usual overidentifying restrictions test can be applied to evaluate the restrictions in \eqref{eqn:R1}. Jointly, Propositions \ref{prop:nec_suff_rank_cond_regime_dependent_irfs}--\ref{prop:asym_dist} provide
the foundation for our approach to the identification and estimation of proxy-SVARs with permanent, nonrecurring breaks in unconditional volatility.
The suggested approach does not necessitate pre-testing proxy strength and exogeneity. In fact, it does not need to rely on weak-instrument robust methods and it does not require imposing proxy exogeneity in estimation.

\subsection{Monte Carlo results}\label{Section_MC_main_test}
The finite-sample performance of the stability restrictions approach is analyzed through Monte Carlo simulations. Specifically, we examine (i) its relative performance compared to using only external instruments, only volatility shifts, or incorrectly imposing constant IRFs across regimes when estimating target IRFs; and (ii) the finite-sample size and power properties of the overidentifying restrictions test in detecting misspecified stability restrictions. Further details on the simulation design, the relative performance measure, and additional results and comments are provided in Section S.5 of the supplementary material. 

Data are generated from a bivariate VAR(1) with one instrument, where a single break in the covariance matrix occurs at mid-sample ($T_B= \lfloor0.5\, T\rfloor$). This break implies a change in the target IRFs, so $\Delta_G \neq0_{3\times3}$. We set $\Psi = I_{3}$ in \eqref{Moment_conditions_one_break_large} for simplicity (this condition is relaxed in Section S.5 of the supplementary material). We explore two main setups: one with a strong instrument (meeting proxy relevance) and one with a local-to-zero instrument. We also consider both exogenous and contaminated instruments, covering all possible proxy properties. 
Importantly, the stability restrictions overidentify the parameters of interest. 

Table~\ref{tab:relmse} compares five models: Model.1 implements the stability restrictions (our benchmark), Model.2 is as Model.1 with instrument exogeneity imposed, while Model.3 uses only volatility shifts. Model.4 uses the proxy but incorrectly assumes constant IRFs across regimes, and Model.5 uses only the proxy, ignoring the volatility shift. We evaluate the performance based on a mean squared errors (MSE) measure designed for IRFs over 25 periods. Results confirm that the stability restrictions approach improves precision and ensures consistency, outperforming methods that rely solely on volatility shifts. If the instrument is genuinely exogenous, enforcing exogeneity can yield gains, but if it is contaminated, leaving it unrestricted is more robust. Local-to-zero instruments limit the advantage over volatility shifts alone, but none of the alternatives surpass the stability restrictions approach even when the instrument is weak and contaminated.

We also examine the overidentifying restrictions test under correct specification, and incorrectly assumed exogeneity. Table~\ref{tab:rejectionfreq} summarizes rejection frequencies at the 5\%  nominal level, for sample sizes $T \in \{250, 500, 1000\}$. When instrument exogeneity is (correctly) not imposed, rejection frequencies are well-controlled regardless of instrument strength.  Conversely, the test displays power against incorrectly imposing exogeneity.

\section{Fiscal Proxies and the Shift from the Great Inflation to the Great Moderation}

\label{Section_empirical_illustration}

In this section, we revisit the seminal US fiscal proxy-SVAR of \cite{MertensRavn2014}, who estimate tax and spending multipliers by combining a (small) VAR for real tax revenues ($TR$), government spending ($GS$), and output ($GDP$), with two external fiscal proxies. We implement the idea that the massive reduction of volatility in macroeconomic variables observed during the transition from the Great Inflation to the Great Moderation macroeconomic regimes led to a change in the dynamic responses of output to the fiscal shocks, rather than solely a change in the variance of these shocks. As robustness check, in Section S.6 of the supplementary material we add consumer price inflation to this baseline model. 

Interestingly, \cite{guay2021identification}, \cite{karamysheva2022we}, and \cite{KewelohKleinPruser2024} have recently contributed to the estimation of U.S. fiscal multipliers on comparable samples using SVARs, leveraging information from higher order moments and non-Gaussian shocks. \cite{Lewis2021} and \cite{FritscheKleinRieth2021} rely on time-varying volatilities. These authors make no use of external instruments and maintain that IRFs do not change across major macroeconomic regimes.

All variables are per capita, deflated by the GDP deflator, and expressed in logarithms. The dataset spans 1950:Q1 to 2006:Q4, for a total of 228 quarterly observations. As in the specifications of \cite{MertensRavn2014} and \cite{CaldaraKamps2017},\ the VAR includes four lags, a linear trend, and a constant.

Let $u_{t}:=(u_{t}^{TR},\, u_{t}^{GS},\, u_{t}^{GDP})^{\prime }$ be the vector of
VAR disturbances, and $\varepsilon _{1,t}:=(\varepsilon
_{t}^{tax},\varepsilon _{t}^{g})^{\prime }$ the vector of (target) fiscal
shocks, $\varepsilon _{2,t}:=\varepsilon _{t}^{y}$ being the (non-target)
output shock. Two fiscal instruments are used for the two fiscal shocks,
collected in the vector $z_{t}:=(z_{t}^{tax},z_{t}^{g})^{\prime }$, $r=k=2$.
Specifically, $z_{t}^{tax}$ represents \cite{MertensRavn2014}'s 
series of unanticipated tax shocks 
identified through a narrative analysis of tax policy decisions, while $z_{t}^{g}$ represents a novel series of unanticipated fiscal spending shocks
introduced in \cite{Angelinietal2023}, to which we refer for details. Interestingly, combining time-varying volatility and independent component analysis with the external instrument approach, \cite{Lewis2021} and \cite{KewelohKleinPruser2024} reject the exogeneity of the tax instrument $z_{t}^{tax}$. The change-point estimator of \cite{Bai2000} detects a shift in VAR parameters, including the error covariance matrix, at $T_B =$1983:Q2. This evidence is consistent with the graphs in Figure S.1 in the supplementary material, which shows a marked reduction in volatility of VAR disturbances since the early 1980s. 
The estimated break point corresponds to the vertical lines in Figure S.1.
The first volatility regime, denoted as the Great Inflation, spans the period from 1950:Q1 to 1983:Q2 and includes 135 quarterly observations. The second volatility regime, denoted as the Great Moderation, covers the
period from 1983:Q3 to 2006:Q4 and includes 93 quarterly observations.

In what follows, our comments focus primarily on the tax shock and tax multipliers for which results appear particularly intriguing.  Detailed comments on the fiscal spending multipliers and comparison with results from other authors are provided in Section S.6 of the supplementary material, which complements our empirical analysis along several dimensions. Table~\ref{tab:peaks} summarizes the main results obtained with the different approaches we discuss below. We quantify estimation uncertainty using 68\% MBB confidence intervals, computed with 4999 bootstrap repetitions.

\paragraph{Ignoring the Volatility Shift.}  We begin by estimating the proxy-SVAR in \eqref{eqn:AC1}--\eqref{eqn:AC2} on the entire sample as in \cite{MertensRavn2014}, imposing the instrument exogeneity condition \eqref{C2}. This yields a peak tax multiplier $\mathcal{M}_{tax}^{peak},$ near 2.6 occurring three quarters after the shock, broadly consistent with their findings (see Table~\ref{tab:peaks}, column (i)). 

A critical parameter influencing the size of
the multiplier is the output elasticity of tax revenues (automatic stabilizer), denoted as $\vartheta_{y}^{tax}$ \citep[see][]{MertensRavn2014,CaldaraKamps2017,Lewis2021}. Our estimate for $\vartheta _{y}^{tax}$, reported in column (i) of Table~\ref{tab:peaks}, is 3.26, with a 68\% confidence interval of (2.48, 5.16). The estimated correlation between the tax instrument $z_{t}^{tax}$ and the recovered tax shock, $\hat{\varepsilon}_{t}^{tax}$, reported in column
(i) of Table~\ref{tab:peaks}, is 27\%, with 68\% confidence interval equal to (12\%, 38\%). The bootstrap-based test of instrument relevance developed in \cite{angelinicavalierefanelli2024} rejects the null hypothesis of strong instruments with a p-value of 0.023; see also \cite{cavaliere2025bootstrap}.   

\paragraph{Volatility Shift with Constant IRFs.} Next, we incorporate the detected volatility break into the analysis keeping the IRFs constant across the Great Inflation and the Great Moderation regimes. The identification of this model, denoted ``proxy-SVAR-H'', depends on the rank condition of Proposition \ref{prop:nec_suff_rank_cond_regime_dependent_irfs}.(ii). The results of this specification are summarized in column (iv) of Table~\ref{tab:peaks} while plots of the implied dynamic fiscal multipliers are in Figure S.2 of the supplementary material. In this case, the estimated tax multiplier plunges below 0.5, and the associated confidence band (-0.93, 1.03) includes zero.  The exogeneity condition \eqref{C2} appears not rejected in this specification, as $corr(z_{t}^{tax},\hat{\varepsilon}_t^{y}) = -17.7\%$ with a wide confidence interval of $(-33.4, 13.9)$. However, the correlation between the proxy and the recovered tax shock is now estimated at the 17\% level with a large confidence interval  (-21.6\%, 26.2\%).  As demonstrated in Corollary 1, if conditions \eqref{C1}--\eqref{C2} hold and instruments are unaffected by the volatility shift, relative responses can be consistently estimated even when the volatility shift is ignored.  The stark contrast from the no-break model ($\mathcal{M}^{peak}_{tax}$ falls from 2.6 to below 0.5) suggests that imposing regime-invariant IRF may conflict with the data. 

\paragraph{Stability Restrictions and Regime-Dependent IRFs.} 
We then implement our stability-restrictions approach, using the external fiscal instruments, incorporating the break in volatility and allowing the IRFs to change before and after 1983:Q2. 
Interestingly, since fiscal multipliers are computed as relative (normalized) responses of output to fiscal shocks (see Section S.6, supplementary material), we forgo placing stability restrictions on $\Psi$, focusing solely on the matrices $G$ and $\Delta _{G}$.
The empirical counterpart of model \eqref{model_changes} is specified as
follows: 
{\setlength{\abovedisplayskip}{5pt}
\setlength{\belowdisplayskip}{5pt}
\begin{align}\nonumber
\begin{pmatrix}
 u_{t}^{TR} \\
 u_{t}^{GS} \\
 u_{t}^{GDP} \\\hline
 z_{t}^{tax} \\
 z_{t}^{g}
\end{pmatrix}
&=
\underset{G}{\underbrace{\left(
\begin{array}{cc|c|cc}
h_{1,1} & h_{1,2} & h_{1,3} & 0 & 0 \\
h_{2,1} & h_{2,2} & 0       & 0 & 0 \\
h_{3,1} & h_{3,2} & h_{3,3} & 0 & 0 \\
\hline
\varphi_{1,1} & \varphi_{1,2} & \upsilon_{tax}^{y} & \omega_{tax} & 0 \\
0             & \varphi_{2,2} & \upsilon_{g}^{y}   & \omega_{g,tax} & \omega_g
\end{array}
\right)}}\,
\underset{\xi_t}{\underbrace{\left(
\begin{array}{l}
\varepsilon_t^{tax} \\
\varepsilon_t^{g} \\ \hline
\varepsilon_t^{y} \\\hline
\zeta_t^{tax} \\
\zeta_t^{g}
\end{array}
\right)}} \\
&\quad+
\underset{\Delta G}{\underbrace{\left(
\begin{array}{cc|c|cc}
\Delta h_{1,1} & \Delta h_{1,2} & \Delta h_{1,3} & 0 & 0 \\
\Delta h_{2,1} & \Delta h_{2,2} & 0              & 0 & 0 \\
0              & \Delta h_{3,2} & \Delta h_{3,3} & 0 & 0 \\
\hline
\Delta\varphi_{1,1} & \Delta\varphi_{1,2} & 0 & 0 & 0 \\
0                   & \Delta\varphi_{2,2} & 0 & 0 & \Delta\omega_g
\end{array}
\right)}}\,
\mathbb{I}(t\ge T_B+1)\;
\underset{\xi_t}{\underbrace{\left(
\begin{array}{l}
\varepsilon_t^{tax} \\
\varepsilon_t^{g} \\\hline
\varepsilon_t^{y} \\\hline
\zeta_t^{tax} \\
\zeta_t^{g}
\end{array}
\right)}}
\label{fiscal_specification_shift}
\end{align}
}
where the zeros in the top-right of $G$ and $\Delta_G$ refer to the impact of measurement errors on $TR_t$, $GS_t$ and $GDP_t$ respectively. 

The structure specified in \eqref{fiscal_specification_shift} satisfies the order condition of Proposition~\ref{prop:nec_suff_rank_cond_regime_dependent_irfs}.(ii) (there are $a+b+c$= 27 free parameters and $%
(n+r)(n+r+1)=30$ moment conditions, implying 3 testable overidentifying restrictions) and is based on several key assumptions. First, the initial two columns of $G$ capture the on-impact effects of fiscal shocks during the Great Inflation, mirroring the single-regime proxy-SVAR structure and leveraging the full-identification power of volatility shifts for both target and non-target shocks. Consistent with \cite{MertensRavn2014} and \cite{CaldaraKamps2017}, we set $h_{2,3} = 0$ so that government spending does not respond to output shocks on impact. We also refrain from forcing exogeneity of the fiscal proxies with respect to output shocks, leaving their potential contamination parameters, $(\upsilon^{y}_{tax}\ ,\, \upsilon^y_g),$ unrestricted. Meanwhile, a zero restriction in the relevance matrix, $\varphi_{2,1}=0,$ is offset by allowing measurement error in the tax proxy to influence the variance of the spending proxy through $\omega_{g,tax}$.

For the Great Moderation, we impose one stability restriction on the tax shock, letting its immediate impact on output remain unchanged ($\Delta_{h_{3,1}} = 0$). This choice is motivated by the stable ratio of $TR_t$ to $GDP_t$ (see Figure S.1 in the supplementary material), implying that the volatility shift affects both series similarly. This is unsurprising, given the tight cyclical connection between real tax revenues and real 
output.  All other on-impact coefficients can shift in the Great Moderation, except the contamination parameters, which remain fixed. However, we allow for a change in the variance of the measurement error linked to the spending proxy. Under this regime-dependent scenario, during the 
Great Inflation (see Table~\ref{tab:peaks}, column (ii)), the tax proxy is relatively weak ($corr(z_{t}^{tax},\hat{\varepsilon}^{tax}_t) = 15.5\%$ with confidence interval $(-8.8\%, 27.8\%)$), yet the estimated peak tax multiplier, $\mathcal{M}^{peak}_{tax}$, is about 1.7 (after eight quarters). We detect, as in \cite{Lewis2021} and \cite{KewelohKleinPruser2024}, a modest negative correlation with the output shock. In fact, $corr(z_t^{tax},\hat{\varepsilon}_t^{y}) = -13.3\%$ with confidence interval $(-25.4\%, -4.5\%)$. By contrast, during the Great Moderation (see Table~\ref{tab:peaks}, column~(iii)), the proxy is stronger ($corr(z_{t}^{tax},\hat{\varepsilon}^{tax}_t) = 45.2\%$ with confidence interval $(21.5\%, 61.2\%)$), but the peak multiplier dips to about 0.5 (on impact) and is estimated less
precisely. The overidentifying restrictions test fails to reject this specification (with a $p$-value of $0.86$, see Table S.5 in the supplementary material for details), indicating that it successfully leverages information from both the external instruments and the volatility shift. Figure~\ref{fig:multipliers} shows markedly different dynamic tax multipliers in the two volatility regimes. 
The underlying IRFs that generate these multipliers are presented and discussed in Section S.6 of the supplementary material. 
Compared with estimating a single-regime proxy-SVAR (black solid line), allowing regime-specific IRFs
produces systematically lower multipliers and tighter confidence intervals. During the Great Inflation, tax cuts exhibit larger and more precisely estimated effects on output, whereas in the Great Moderation, tax shocks appear less
potent.  These patterns imply that ignoring changes in volatility, evolving proxy strength, and even mild contamination can bias full-sample multipliers upward -near 3 in \citealp{MertensRavn2014} or 2.6 in our own estimation. By contrast, modeling the volatility shift reveals that exogenous tax shocks affect output quite differently across regimes. This is particularly evident from the graphs of IRFs plotted in Figure S.3 of the supplementary material, Section S.6. Overall, our findings underscore that shifts in volatility, changes in proxy relevance, and the relaxation of the proxy exogeneity assumption (while retaining the economic information provided by the narrative instrument) can significantly affect inference on fiscal multipliers. It is worth briefly comparing our dynamic tax multipliers in Figure~\ref{fig:multipliers} with those plotted in Figure 8 of \citealp{MertensRavn2014} (right panel), which were obtained by simply splitting the estimation sample before and after 1980. We observe that in the right panel of Figure 8 of \citealp{MertensRavn2014}, the difference between tax multipliers in the two sub-samples appears minor, especially on-impact. Our empirical evidence stands in stark contrast to their findings, suggesting their results likely depend on the narrative tax instrument being potentially weak and contaminated. Augmenting the baseline model with consumer‑price inflation (see Section S.6 in the supplementary material) leaves our central results intact. 

\section{Concluding remarks}

\label{Section_concluding_remarks}
 Permanent, exogenous volatility shifts in proxy-SVARs affect estimation consistency if not properly addressed. However, when stability restrictions are employed to accurately incorporate volatility shifts into the analysis, they not only restore estimation consistency but also provide a framework where even statistically invalid instruments may positively contribute to the identification. Our empirical illustration based on US quarterly data, demonstrates  that the narrative proxy used for the tax shock, despite being potentially contaminated by the output shock and exhibiting different degrees of relevance across the Great Inflation and Great Moderation periods, still contributes to revealing the role of tax policy in stabilizing business cycle fluctuations.

{\singlespacing
\vspace{-10pt}
\subsection*{Acknowledgements}
We gratefully acknowledge financial support from MIUR (PRIN 2022, Financed by European Union - Next Generation EU, Mission 4 Component 2, Grant 20229PFAX5\_002, and Grant 2022H2STF2) and the University of Bologna (RFO grants). Luca Neri acknowledges support from the Carlo Giannini Association and the EU Horizon Europe programme (Marie Skłodowska-Curie grant No.~101212449, MacHete).
A list of seminars/conferences where this paper has been presented and participants to whom we wish to thank 
is reported in the extended acknowledgments section of the supplementary material.
\vspace{-10pt}

\bibliographystyle{apalike}   
\bibliography{phs}

\begin{table}[H]
\caption{Relative performance (MSE) of estimators of the target IRFs. }
    \resizebox{\columnwidth}{!}{
            \begin{tabular}{lccccccccccc}
                    \hline\hline
                    Sample size: $T=500$&  \multicolumn{11}{c}{$corr(z_t, \varepsilon_{2t})$ } \\
                     & \multicolumn{2}{c}{0.00}& &\multicolumn{2}{c}{0.05}& &\multicolumn{2}{c}{0.15}& &\multicolumn{2}{c}{0.25}
                    \\\cline{2-3}\cline{5-6} \cline{8-9} \cline{11-12}
                    & $IRF_{1,1}$ & $IRF_{2,1}$& & $IRF_{1,1}$ & $IRF_{2,1}$& &$IRF_{1,1}$ & $IRF_{2,1}$& &$IRF_{1,1}$ & $IRF_{2,1}$ \\\hline
                     \textit{Panel a) Strong proxy}& & & &&&&&  &&&\\  \cline{1-1}
                      Model.1  & 1.00& 1.00& & 1.00& 1.00& & 1.00& 1.00& & 1.00& 1.00 \\ 
                     Model.2  & 0.95& 0.96& & 1.01& 1.02& &1.33& 1.36& &1.95& 2.04 \\ 
                     Model.3 & 11.75& 7.11& &13.34& 9.03& &16.69& 13.36& &18.36& 15.25 \\ 
                      Model.4 & 5.87& 4.10& &5.89& 4.19& &5.94& 4.39& &6.05& 4.54 \\ 
                     Model.5  & 4.62& 2.71& &5.41& 3.37& &7.15& 5.24& &8.51& 7.01 \\ 
                    \hline
                    \textit{ Panel b) Local-to-zero proxy}& & &&&&&& &&&  \\ \cline{1-1}
                      Model.1  & 1.00& 1.00& &1.00& 1.00& &1.00& 1.00& &1.00& 1.00 \\ 
                     Model.2  & 1.00& 1.00& &1.00& 1.01& &1.02& 1.03& &1.07& 1.07 \\
                     Model.3  & 1.00& 1.00& & 1.00& 1.00& &1.00& 0.99& & 1.00& 0.99 \\ 
                     Model.4  & 5.80& 3.86& & 5.95& 3.93& & 5.69& 3.90& &5.72& 3.94 \\ 
                      Model.5    & 13.27& 11.92& &11.12& 10.77& & 7.26& 7.86& &5.85& 7.05 \\  
                    \hline\hline
            \end{tabular}}
{ {
                    \footnotesize Notes: Results are based on $N=$10,000 Monte Carlo simulations, see Section
                    S.5 for details on the design. Model.1 denotes results obtained by the
                    stability restrictions approach discussed in the paper. Model.2 is the
                    same as Model.1 with the contamination parameters  in $\Upsilon $ and $\Delta
                    _{\Upsilon }$ set to zero, i.e., imposing proxy exogeneity. Model.3
                    denotes results obtained by the change in volatility approach alone, i.e.,
                    without including the instrument. Model.4 denotes results obtained by the
                    proxy-SVAR-H approach, see Proposition \ref{prop:nec_suff_rank_cond_regime_dependent_irfs}.(iii), i.e., assuming that the target IRFs
                    remain constant across the two volatility regimes. Model.5 denotes results                     obtained by the external instrument alone, i.e., ignoring the volatility break. Numbers in the table correspond to measures of relative                   performance in the estimation of target IRFs based on Mean Squared Error
                    (MSE), as discussed in Section S.5. Model.1 is used as a benchmark in the
                    comparison; thus, relative performance measures are set to 1 for this model.}}
            
            \label{tab:relmse}
\end{table}
\begin{table}[H]
            \caption{Rejection frequencies of the overidentifying restrictions test
            (5\% nominal).} \footnotesize
    \resizebox{\columnwidth}{!}{
            \begin{tabular}{llccccccccc}
                    \hline\hline
                    & & \multicolumn{9}{c}{$corr(z_t, \varepsilon_{2t} )$} \\ 
                    & & \multicolumn{4}{c}{$\Upsilon$ is set to $0$} & & \multicolumn{4}{c}{$\Upsilon$ is unrestricted}  \\ \cline{3-6} \cline{8-11}
                    & &$0.00$&$0.05$&$0.15$&$0.25$& & $0.00$&$0.05$&$0.15$&$0.25$ \\\hline
                    Sample size & Relevance &\multicolumn{9}{c}{Rejection frequency ($5\%$)} \\\hline
                    $T=250$ & Strong & 4.06& 7.64& 40.84& 88.30& &4.83& 4.63& 4.42& 4.43 \\ 
                    & Local-to-zero & 4.28& 8.22& 45.73& 91.83& &4.22& 4.72& 4.34& 4.67 \\ 
                    $T=500$ &Strong & 4.68& 12.03& 75.26& 99.80& &4.87& 4.73& 4.55& 4.49 \\ 
                    & Local-to-zero & 4.34& 13.26& 80.48& 99.92& &4.67& 4.93& 4.79& 5.10 \\ 		
                    $T=1000$ &Strong & 4.22& 21.40& 97.64& 100.00& & 5.21& 5.04& 4.57& 5.09 \\ 
                    & Local-to-zero & 4.64& 22.62& 98.60& 100.00& &4.74& 4.73& 5.04& 5.06 \\ 
                    \hline\hline
                \end{tabular}
        }
{  {Notes: Rejection frequencies are computed across $N=$10,000 Monte Carlo
                    simulations, see Section S.5 for details on the design. Estimates of
                    proxy-SVAR\ parameters are obtained by the CMD approach discussed in Section
                    \ref{sec:estimation}.}}
                    \label{tab:rejectionfreq}
\end{table}	
\vspace{-10pt}
\begin{table}[H]
	\caption{Estimated peak multipliers and elasticities with 68\% MBB confidence intervals (in  parentheses).  Volatility shift date $T_{B}$ = 1983:Q2.}
     \resizebox{\columnwidth}{!}{
	\begin{tabular}{lcccc}
	\hline \hline
	& \tiny(i) &  \tiny(ii) &  \tiny(iii) &  \tiny(iv) \\
	& $\underset{1950:Q1 - 2006:Q4}{\text{Proxy-SVAR}}$ & $\underset{1950:Q1 - 1983:Q2}{1^{st}\text{ vol. regime}}$ & $\underset{1983:Q3 - 2006:Q4}{2^{nd}\text{ vol. regime}}$ &  $\underset{\text{break at } 1983:Q2}{\text{ Proxy-SVAR-H}}$\\
	\\
	\hline
	$\vartheta_{y}^{tax}$ & $\underset{(2.483, 5.163)}{3.256}$ &$\underset{(1.388, 2.219)}{1.924}$ &$\underset{(0.702, 4.801)}{2.812}$ &$\underset{(-11.577, 7.023)}{1.680}$ \\ 
$\mathcal{M}^{peak}_{tax}$ & $\underset{(0.747, 5.843)}{2.625(3)}$ &$\underset{(0.965, 2.635)}{1.726(8)}$ &$\underset{(-0.288, 1.077)}{0.535(0)}$ &$\underset{(-0.929, 1.034)}{0.459(9)}$ \\
relevance$_{tax}$ (\%) 
&$\underset{(11.8, 37.6)}{27.1}$
& $\underset{(-8.8, 27.8)}{15.5}$
& $\underset{(21.5, 61.2)}{45.2}$
& $\underset{(-21.6, 26.2)}{17.0}$ 
\\ 
contamination$_{tax}$ (\%) & - &$\underset{(-25.4, -4.5)}{-13.3}$ 
&$\underset{(-22.6, -2.6)}{-11.9}$ 
&$\underset{(-33.4, 13.9)}{-17.7}$ 
\\
$\vartheta_{y}^{g}$ & $\underset{(-0.031, 0.037)}{-0.005}$ &$\underset{(-0.008, 0.052)}{0.027}$ &$\underset{(-0.057, -0.001)}{-0.026}$ &$\underset{(-0.035, 0.025)}{-0.031}$ \\ 
$\mathcal{M}^{peak}_{g}$ & $\underset{(0.769, 1.914)}{1.671(4)}$ &$\underset{(1.176, 2.502)}{2.405(4)}$ &$\underset{(1.275, 2.706)}{2.028(2)}$ &$\underset{(1.460, 1.901)}{1.514(5)}$ \\ 
relevance$_{g}$ (\%) & $\underset{(96.4, 98.0)}{96.5}$
&$\underset{(95.8, 97.8)}{96.0}$
&$\underset{(97.2, 99.0)}{98.0}$ 
&$\underset{(96.1, 98.2)}{97.0}$ 
\\ 
contamination$_{g}$ (\%) & - & $\underset{(-0.1, 0.7)}{0.5}$
&$\underset{(-0.1, 1.5)}{0.9}$ 
& $\underset{(-0.2, 3.7)}{2.0}$ 
\\
					\hline \hline
				\end{tabular}}
		{ \footnotesize Notes: All columns use external instruments $z_t:=(z_t^{tax}, z_{t}^g)'$. Column (i) presents estimates from the proxy-SVAR approach using the full sample (1950:Q1–2006:Q4)  without accounting for volatility shifts. Column (ii) shows estimates for the first volatility regime (1950:Q1–1983:Q2). Column (iii) provides estimates for the second volatility regime (1983:Q3–2006:Q4)s. Column (iv) presents estimates from the proxy-SVAR-H approach, assuming constant IRFs across volatility regimes. $\vartheta_{y}^{tax}$ and $\vartheta_{y}^{g}$ are the elasticities of tax revenue and fiscal spending to output, respectively. $\mathcal{M}^{peak}{tax}$ and $\mathcal{M}^{peak}{g}$ denote peak multipliers. ``Relevance${(\cdot)}$ (\%)\'' denotes the correlation between the instrument $z{t}^{(\cdot)}$ and the estimated shock $\hat{\varepsilon}t^{(\cdot)}$. ``Contamination${(\cdot)}$ (\%)'' refers to the correlation between the instrument $z_{t}^{(\cdot)}$ and the non-target output shock $\hat{\varepsilon}_t^{y}$.}	
\label{tab:peaks}
	\end{table}
\begin{figure}[H]
\caption{Estimated dynamic fiscal multipliers with 68\% MBB (pointwise)\
confidence intervals.}
\begin{center}
\begin{tabular}{c}
\includegraphics[width=12cm,keepaspectratio]{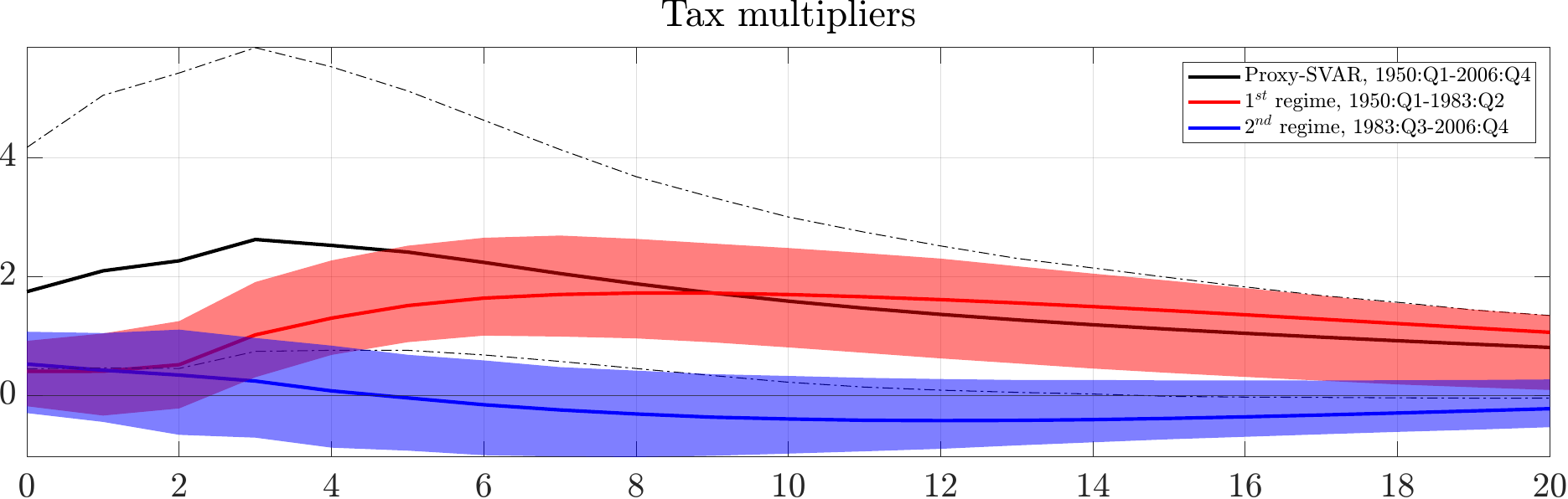}   \\
\includegraphics[width=12cm,keepaspectratio]{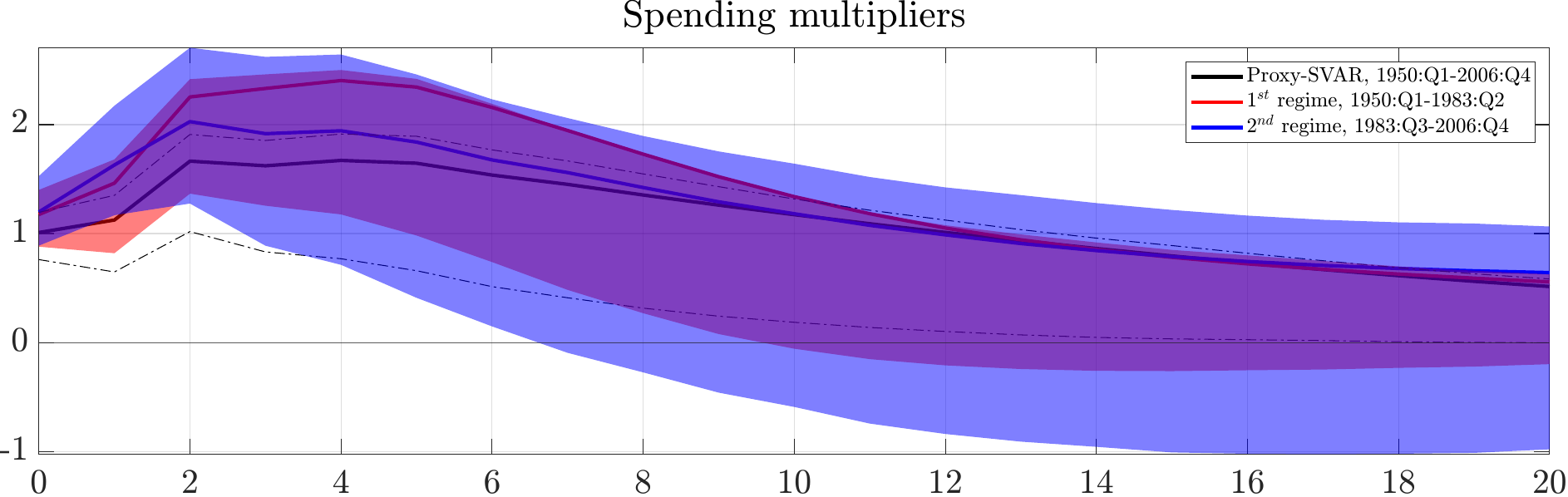}   \\
\end{tabular}
\label{fig:multipliers}
\end{center}\vspace{-8pt}
{ \footnotesize Notes: Tax multipliers are in the
upper panel; fiscal spending multipliers in the lower panel. Black solid
lines refer to multipliers estimated on the whole sample 1950:Q1--2006:Q4,
without accounting for the detected shift in volatility; dotted thin black lines are
the associated 68\% MBB confidence intervals. Red solid line refer to
multipliers estimated on the first volatility regime 1950:Q1--1983:Q2 (Great
Inflation); red shaded areas are the associated 68\% MBB confidence
intervals. Blue solid lines refer to multipliers estimated on the second
volatility regime, 1983:Q3--2006:Q4 (Great Moderation); blue shaded areas
are the associated 68\% MBB confidence intervals.}
\label{fig:multipliers}
\end{figure}

\end{document}